\documentclass[12pt]{article}
\usepackage{graphicx}
\usepackage{natbib}
\usepackage{url}

\usepackage[
nodocdata=true,
noeditdata=true,
noproducerdata=true,
noptexdata=true,
nopdftrailerid=true
]{pdfprivacy}

\newcommand{\blind}{0}

\RequirePackage{luatex85}
\usepackage{authblk}  
\usepackage{booktabs} 
\usepackage{amsmath}
\usepackage{amssymb}
\usepackage{mathtools}
\usepackage{amsthm}
\usepackage{adjustbox}
\usepackage{bm}
\usepackage{float}
\usepackage{pdflscape}
\usepackage{hyperref}
\usepackage[capitalize,noabbrev]{cleveref}
\usepackage{xr}
\usepackage{soul}

\usepackage{caption}
\newcommand{\Reals}{\mathbb{R}}
\newcommand{\RealsP}{{\Reals}^{+}}

\DeclareMathOperator*{\argmax}{arg\,max}

\newcommand{\norm}[1]{\left\lVert#1\right\rVert}

\newcommand{\E}{\operatorname{E}}
\newcommand{\V}{\operatorname{V}}
\newcommand{\opSE}{\operatorname{SE}}

\newcommand{\EE}[1]{\E\left[#1\right]}
\newcommand{\VV}[1]{\V\left[#1\right]}

\newcommand{\iid}{\smash{\overset{\text{\tiny iid}}{\sim}}}

\newcommand{\viGP}{{\textsc{viGP}}}

\newcommand{\fiGP}{{\textsc{fiGP}}}

\newcommand{\fiGPADE}{{\textsc{fiGP-ADE}}}

\newcommand{\PFI}{\textsc{PFI}}
\newcommand{\PFDI}{\textsc{PFDI}}

\newcommand{\predmean}{\hat{\mathbf{m}}^{y}_*}
\newcommand{\predvar}{\hat{\mathbf{S}}^{y}_*}

\graphicspath{{inc/}}

\begin{document}

\def\spacingset#1{\renewcommand{\baselinestretch}%
{#1}\small\normalsize} \spacingset{1}


\author[1]{Luis Damiano}
\author[2]{Margaret Johnson}
\author[2]{Joaquim Teixeira}
\author[3]{Max D. Morris}
\author[1]{Jarad Niemi}
\affil[1]{\small Department of Statistics, Iowa State University, Ames, IA, US}
\affil[2]{\small Jet Propulsion Laboratory, California Institute of Technology,
  Pasadena, CA, US}
\affil[3]{\small Departments of Statistics, and Industrial and Manufacturing
Systems Engineering, Iowa State University, Ames, IA, US}

\if0\blind{}
{
  \title{\bf
    Automatic Dynamic Relevance Determination \\
    for Gaussian process regression \\
    with high-dimensional functional inputs
  }
  \maketitle
} \fi

\if1\blind{}
{
  \bigskip
  \bigskip
  \bigskip
  \begin{center}
    { \LARGE\bf
      Automatic Dynamic Relevance Determination \\
      for Gaussian process regression \\
      with high-dimensional functional inputs }
  \end{center}
  \medskip
} \fi

\medskip
\begin{abstract}
  In the context of Gaussian process regression with functional
  inputs, it is common to treat the input as a vector. The parameter
  space becomes prohibitively complex as the number of functional points
  increases, effectively becoming a hindrance for automatic relevance
  determination in high-dimensional problems.  Generalizing a framework
  for time-varying inputs, we introduce the asymmetric Laplace
  functional weight (ALF): a flexible, parametric function that drives
  predictive relevance over the index space. Automatic
  dynamic relevance determination (ADRD) is achieved with three
  unknowns per input variable and enforces smoothness over the index space.
  Additionally, we discuss a screening technique to assess under
  complete absence of prior and model information whether ADRD is
  reasonably consistent with the data. Such tool may serve for
  exploratory analyses and model diagnostics.
  ADRD is applied to remote sensing data and predictions are generated in
  response to atmospheric functional inputs. Fully Bayesian estimation is carried
  out to identify relevant regions of the functional input space. Validation is
  performed to benchmark against traditional vector-input model specifications. We
  find that ADRD outperforms models with input dimension reduction via functional
  principal component analysis.
  Furthermore, the predictive power is comparable to high-dimensional models,
in terms of both mean prediction and uncertainty, with 10 times fewer tuning
parameters. Enforcing smoothness on the predictive relevance profile rules out
erratic patterns associated with vector-input models.
\end{abstract}

\noindent%
\vfill

\newpage
\spacingset{2}
\section{Introduction}

The study of uncertainty in complex physical systems commonly relies
on the evaluation of computationally expensive computer models.  Many
science and engineering computer models approximate systems that
include functional inputs, i.e., input quantities varying over a
continuum typically modeled as function of some index. Dynamical
systems are a widely popular sub-group of this family whose main
characteristic is that the input, and possibly the output, is a
function of time. In other settings, the index space could represent
space or wavelengths.

Gaussian processes are a common choice among the many available
statistical and machine learning models for computer model emulation,
with early work such as
\citet{sacks1989,sacks1989a,currin1991,ohagan1992,koehler1996,jones1998,kennedy2001}.
A modern, general treatment can be found in
\citet{santner2018,gramacy2020}.
However, the literature mainly emphasizes the emulation of dynamical systems via
Gaussian processes with scalar, vector scalar-summary, or possibly pre-processed
vector inputs in relation to either a scalar output or a pre-processed
structured output modeled as separable
vector~\citep{campbell2006,bayarri2007a,higdon2008}.

A simple modeling strategy is to consider the functional input not as
functional, but an ensemble of possibly many scalar
values and a specific correlation structure~\citep{iooss2009}. Although
straightforward, discretization leads to a potentially unlimited number of
feature variables used to predict a finite number of scalar outputs, thus
increasing model training difficulty and the risk of overfitting.
Functional input pre-processing often involves the application of
techniques for scaling, registration, warping, decorrelation,
dimensionality reduction, and regularization. Frequently
used ones are a-priori basis expansion, e.g.,
splines~\citep{betancourt2020,betancourt2020a}, principal component
analysis~\citep{nanty2016}, functional principal component
analysis~\citep{wang2017,wang2019}, among other basis
functions~\citep{tan2019,li2021}.  Such decompositions introduce
undesirable side effects such as the laborious interpretation of
non-physical variables, and the lack of direct use of the information
contained in the input structure, e.g., the notions of order and
closeness incorporated in the index space.  These also
struggle with the modeling and computational challenges
associated with high-dimensional input vectors, principally a large
number of unknowns and the risk of overfitting.

Whatever the approach, the complexity of present applications
routinely warrants the use of regularization. Automatic relevance
determination priors~\citep{neal1996} are prevalent in the world of
Gaussian processes. Each input dimension, be a measurement from a
functional profile or a basis coefficient, is given a different
length-scale parameter that allows the latent function to vary at
different speeds with respect to different inputs. This introduces
multiple regularization as the marginal likelihood would favor
solutions with large length-scales for those inputs along which the
latent function is flat, a mechanism for Bayesian Occam's razor put in
place to prevent significant overfitting by pruning high-dimensional
inputs toward sparse representations~\citep{mackay1996,wipf2007}.  In
this context, relevance is defined as the inverse of the length-scale
even though this quantity is also affected by other intrinsic
characteristics of the data such as the output responding linearly or
non-linearly to changes in the input variable \citep{piironen2016}.
These length-scale parameters are sometimes known as correlation
lengths~\citep{santner2018} or ranges~\citep{cressie1993}, and their inverse as
roughness~\citep{kennedy2001}.









%

Relatively less work has been done to incorporate the input functional form into
Gaussian process regression models, and the capabilities for automatic relevance
determination in the developed methods are rather limited.
Time-indexed input-output pairs for dynamical systems
were discussed by~\citet{morris2012}, who proposed a half-normal
weight function for time-indexed inputs where physical knowledge
suggests a reversion to a neutral state or a reaction slowdown
relative to some fixed time point. In this work, a point Bayesian
estimate was produced for a computer experiment with five runs and a
profile with 13 time steps.
Later on,~\citet{muehlenstaedt2017} presented two approaches for the
$L^2$ norm for functional inputs via projection-based methodologies
coupled with weights assigned to the basis functions.
Such weights remain constant over the index space and originate in the
discretization of the linear and beta
distributions.~\citet{kuttubekova2019} built a parametric weight via
trigonometric basis functions of the functional index and predicted
soil loss from daily precipitation and hill slope profiles with a
one-harmonic Fourier expansion.  Alternatively, \citet{chen2021}
introduced kriging for functional inputs with functional weights in the
spectral domain associating relevance quantities to each functional
input frequency component rather than to the input measurements
themselves.

With the ultimate goal of pushing the emulation of computer models
with functional inputs closer to a hands-off process, we expand on
these ideas and introduce a new functional weight parametric form for
automatic dynamic relevance determination (ADRD). The asymmetric
Laplace functional weight (\textsc{ALF}) learns from the data how the input
predictive relevance
varies across the input index space.  Building
off the notion that a predictive model may assign different weights
across this index space, we set up a parametric
weight function to enforce smoothness on relevance over the index
space and allow the model to learn that some index subspaces are more
relevant than others.

We describe the \textsc{ALF} methodology in \cref{sec:methods}, including
procedures for fully-Bayesian estimation of the model parameters,
model validation, and dynamic relevance screening in
\cref{sec:estimation,sec:validation,sec:feature-importance}.
In \cref{sec:application},
we introduce a case study concerning the emulation of a computer
model with functional inputs. We compare the posterior estimates for
the automatic relevance determination parameters and out-of-sample
predictive power across several plausible models and sets of input
variables. Finally, we present in \cref{sec:discussion} potential
improvements on our current work and delineate our future line of
work.


\section{Methods}\label{sec:methods}

Consider an experiment with a functional input $X(t) \in \mathcal{X}$
indexed by a continuous index $t \in \mathcal{T}$ and a scalar output
$y \in \mathcal{Y}$. We model the unknown input-output relationship
$f: \mathcal{X} \to \mathcal{Y}$ via a Gaussian process with mean
function $m_y: \mathcal{X} \to \mathcal{Y}$ and a positive definite
covariance function $s_f: \mathcal{X}^2 \to \RealsP$.
Consider the squared exponential covariance function with
homogeneous independent noise
$s_f(X_i, X_j) = \sigma_{f}^2 \ \exp
\left\{ - 0.5 \ d(X_i, X_j) \right\}$,
$s_y(X_i, X_j) = s_f(X_i, X_j) + I_{i, j} \
\sigma_{\varepsilon}^2, $
where $\sigma_{f}^2 \in \RealsP$ is the
output signal variance, $\sigma_{\varepsilon}^2 \in \RealsP$ is the
output noise variance, $I_{i,j}$ is an indicator function for $i = j$,
and $i, j = 1, \dots, N \in \mathbb{N}$ index the functional
profiles. The function $d: \mathcal{X}^2 \to \RealsP$ quantifies the
distance in any functional input pair. If the notion of a functional
input is completely simplified to a $K-$dimensional vector input
coupled with an automatic relevance determination (ARD) hierarchical
prior for separable length-scale parameters, we have
$d_{\textsc{ARD}}(X_i, X_j) = \sum_{k = 1}^{K}{(x_{i,k} - x_{j, k})}^2
/ \sigma_{x_k}^2$ for $\sigma_{x_k}^2 > 0 \ \forall \ k = 1, \dots, K
\in \mathbb{N}$.  Alternatively, when the functional input is
pre-processed via functional principal component
analysis~\citep{ramsay2005}, then
$d_{\textsc{PC}}(X_i, X_j) = \sum_{{\tilde{k}} = 1}^{{\tilde{K}}}{
  {({\tilde{x}}_{i,{\tilde{k}}} - {\tilde{x}}_{j, {\tilde{k}}})}^2 } /
\sigma^2_{{\tilde{x}}_{{\tilde{k}}}}$ for
$\sigma_{{\tilde{x}}_{\tilde{k}}}^2 > 0 \ \forall \ {\tilde{k}} = 1,
\dots, {\tilde{K}} \le K$, where ${\tilde{x}}_{i, {\tilde{k}}}$ is the
${\tilde{k}}$-th score for the $i$-th profile and ${\tilde{K}}$ is the
total number of principal components retained in the model.

Recognizing that the profile measurements $\mathbf{x}_i = {\left\{x_{i,
k}\right\}}_{k}$ are a finite representation of an
infinite-dimensional function, we can quantify the distance between
any two profile inputs via the weighted functional norm, i.e.,
\begin{equation}\label{eq:functional-norm}
d_{\omega}(X_i, X_j) =
\phi^{-2} \int_{\mathcal{T}} \omega(t) {\left(X_i(t) -
    X_j(t)\right)}^2 \, \mathrm{d}t
\end{equation}
where $\phi > 0$ and $\omega: \mathcal{T} \to
\RealsP$ is a weight function driving the predictive relevance of the
input on the output over the index space.
A constraint is needed for $\omega(t)$ and $\phi$ to be
identifiable, e.g., $\omega(t) = 1$ for some $t \in \mathcal{T}$ or
$\int_{\mathcal{T}} \, \omega(t) \, \mathrm{d}t = 1$.
The functional form of
$\omega(t)$ is a modeling decision that opens up the possibility for
domain-specific knowledge about the physical system; in truly complex
settings, its specification might require a fully data-driven
approach.

\begin{figure}
  \centering
    \includegraphics[width=.79\textwidth]{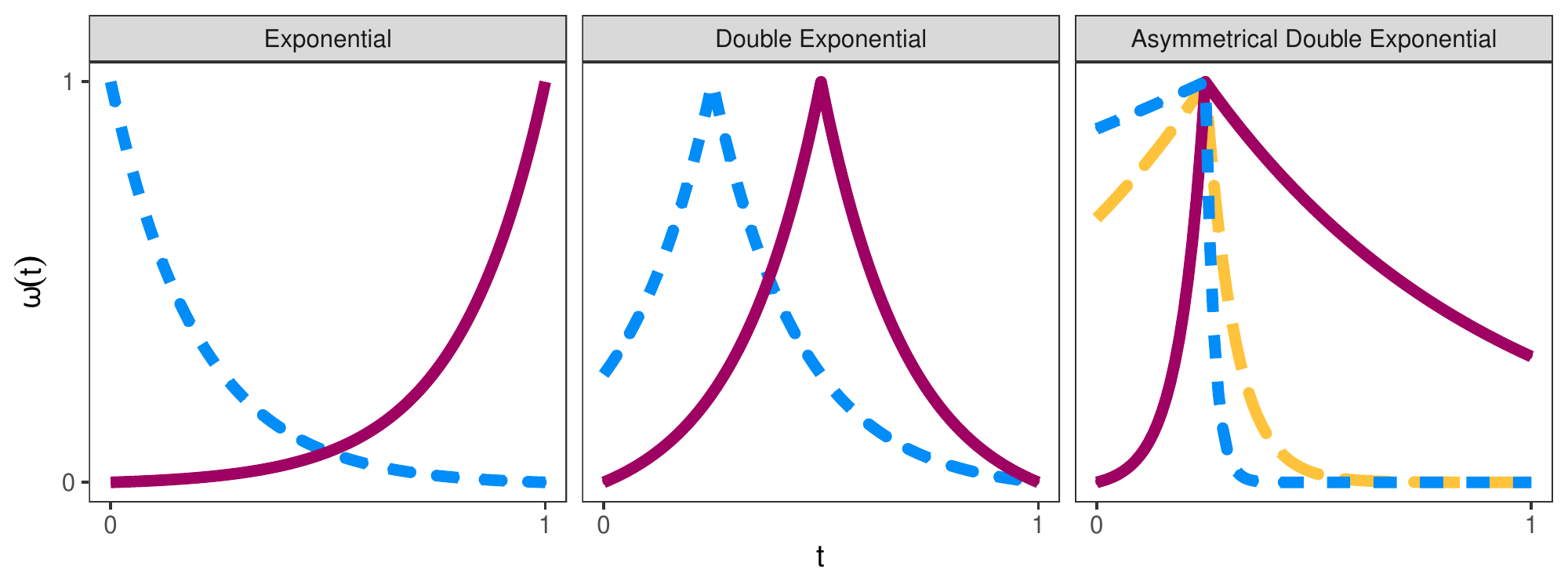}
    \caption[True weight functions]{Some selected \textsc{ALF} configurations.
      Colors do not indicate grouping across facets.}%
  \label{fig:alf-weight-plot}
\end{figure}

In practice, even if subsystems are well understood and documented,
computer models consolidate a large number of interrelationships that
may hinder prior relevance elicitation. In an effort to provide the
weight function with enough flexibility for a more data-driven
approach, we introduce the asymmetric Laplace functional weight (\textsc{ALF}),
\begin{align}
  \label{eq:weight-alf}
  \omega(t)
  &= \text{exp}\left(-(t - \tau) \lambda \kappa^s s\right)
  =
    \begin{cases}
      \text{exp}\left(- \lambda_1 \ \lvert t - \tau \rvert
      \right) & \text{for } t \le \tau \\
      \text{exp}\left(- \lambda_2 \ \lvert t - \tau \rvert
      \right) & \text{for } t > \tau \\
    \end{cases}
\end{align}
where we assume $\mathcal{T} = [0, 1]$ without loss of generality, $s
= \text{sign}(t - \tau) \in \{-1, 1\}$, $\lambda > 0$ is the rate,
$\tau \in \mathcal{T}$ is the location, and $\kappa > 0$ is the
asymmetry coefficient.
An alternative parametrization has $\lambda_1 =
\lambda \kappa^{-1}$ and $\lambda_2 = \lambda \kappa$ for the left and
right exponential decay rates around $\tau$.
\Cref{fig:alf-weight-plot} shows that \textsc{ALF} encompasses a wide family of
functions such as the exponential decay ($\kappa = 1, \tau = 0$), exponential
growth ($\kappa = 1, \tau = 1$), symmetric exponential with predictive relevance
peak at $\tau$ ($\kappa = 1$), near-zero weight before or after the location
$\tau$ ($\lambda_1 \to \infty$ or $\lambda_2 \to \infty$), and constant weight
before and after the location $\tau$ ($\lambda_1 \to 0$ or $\lambda_2 \to 0$).

This parametric function provides enough degrees of freedom for a
rather wide range of unimodal patterns in predictive relevance with at
most three parameters, attempting to find a balance in learnability
and parameter space complexity compared to an \textsc{ARD} model with
a typically large number of tuning unknowns. \textsc{ALF} offers a method to
grow the resolution of the input profiles without compromising the
parameter space complexity, effectively circumventing pre-processing
decisions for input dimension reduction.  Equally appealing, it
enforces smoothness on predictive relevance over the index space
ruling out erratic patterns for relevance in the fully-free
\textsc{ARD} learnable space, e.g., $X(t_{k_1})$, $X(t_{k_2})$ and
$X(t_{k_3})$ having disparate weights despite $t_{k_1} < t_{k_2} <
t_{k_3}$ being close on the index space for the scale of the physical
problem at hand.  Besides making high-dimensional inputs more
manageable, in situations where there is at least subsidiary interest
in interpretable predictive modeling, \textsc{ALF} offers some
insight in the input-output relationship by establishing an explicit
link between the functional input index space and the output
correlation. Such insight enters the predictive modeling feedback loop
and may also foster the understanding of the underlying physical
model. Overall, \textsc{ALF} attempts to produce a simplified representation
that addresses interpretation, smoothness differentiation, and
parsimony in relevance determination.

\subsection{Estimation}\label{sec:estimation}

Consider the functional-input Gaussian process models (\textsc{fiGP})
with $d_{\omega}(\cdot)$ and the following configurations:
$\kappa = 1$ and $\tau = 0$ decreasing exponential (\textsc{fiGP-Edn}),
$\kappa = 1$ symmetric double exponential (\textsc{fiGP-SDE}),
and asymmetric double exponential with all three parameters free (\textsc{fiGP-ADE}).
We use the umbrella term \textsc{ALF} to refer to these three models
altogether since these are all parametrizations of
\cref{eq:weight-alf}. We denote the parameter vector by $\bm{\theta} =
(\bm{\theta}_{d}, \sigma_f^2, \sigma_{\varepsilon}^2)$, where
$\bm{\theta}_{d}$ encompasses the unknowns for a specific choice of
$d(X_i, X_j)$, e.g., $\bm{\theta}_{d} = (\phi, \lambda, \kappa, \tau)$ for
\cref{eq:functional-norm,eq:weight-alf},
while the signal and noise variances are present in
all the considered models.
We generally recommend the following independent weakly informative
priors for cases where there is no domain-specific information about
the relevance profile:
$\phi \sim \textsc{InvGamma}(\cdot, \cdot)$,
$\tau \sim \textsc{Beta}(\cdot, \cdot)$,
$\lambda \sim \textsc{N}^{+}(\cdot, \cdot)$,
$\log(\kappa) \sim \textsc{N}(\cdot, \cdot)$ for the \fiGP{} parameters, and
$\sigma_f, \sigma_{\varepsilon} \iid \textsc{N}^{+}(\cdot, \cdot)$
for the standard deviation parameters.
The choice
of prior for $\phi$ is motivated by the fact that values extremely
close to zero or large may lead to numerical issues due to posterior
density flattening.
The integral in \cref{eq:functional-norm} is
computed via the trapezoidal approximation,
i.e., $\smash{d_{\omega}(X_i, X_j) \approx \
    \phi^{-2}
    \sum_{k = 2}^{K} {
      \left(t_{k} - t_{k - 1}\right)
      (\Delta_{i, j, k} + \Delta_{i, j, k - 1})/2
    }}$ where
  $\Delta_{i, j, k} =
    \omega(t_{k}) {\left(x_{i, k} - x_{j, k}\right)}^2$.

We place a Gaussian process prior on the unknown function, i.e.,
$\mathbf{f}(\mathbf{X}) \sim \mathcal{N}( \mathbf{m}_y,
\mathbf{S}_y)$, where $\mathbf{f}(\mathbf{X})$ is a vector of function
values, $\mathbf{X}$ is the training input matrix, $\mathbf{m}_y =
m_y(\mathbf{X})$ is the mean vector with elements
${(\mathbf{m}_y)}_{i} = m_y(\mathbf{x}_i)$, and $\mathbf{S}_y =
s_y(\mathbf{X})$ is the covariance matrix with elements
${(\mathbf{S}_y)}_{i, j} = s_y(X_i, X_j | {\bm{\theta}})$. Denoting
the training output vector $\mathbf{y}$ and the covariance function
parameter joint prior $p(\bm{\theta})$, the output marginal log likelihood
and the parameter posterior density are given
in~\cref{eq:margina-likelihood,eq:parameter-posterior} respectively,
\begin{align}
  \label{eq:margina-likelihood}
  \log p(\mathbf{y} | \mathbf{X}, \bm{\theta})
  =& -\frac{1}{2}
     {(\mathbf{y} - \mathbf{m}_y)}^\top
     {\mathbf{S}_y}^{-1}
     {(\mathbf{y} - \mathbf{m}_y)}
     -\frac{1}{2}
     \log | \mathbf{S}_y |
     - \frac{n}{2} \log 2\pi \\
  \label{eq:parameter-posterior}
  \log p(\bm{\theta} | \mathbf{y}, \mathbf{X})
  \propto&
	   \log p(\mathbf{y} | \mathbf{X}, \bm{\theta}) +
	   \log p(\bm{\theta}).
\end{align}

We perform fully Bayesian inference on the unknown quantities. The log
posterior~\eqref{eq:parameter-posterior} is evaluated at random
locations of the parameter space.
A number of candidate parameter vectors with highest posterior density are
selected and an optimization is initialized at each candidate.
The
optimal values with highest posterior density are used to initialize
one MCMC chain~\citep{raftery1992} with $M \in \mathbb{N}$ post warm-up
samples generated via the NUTS algorithm~\citep{hoffman2014}. Lack of
non-convergence is diagnosed by the absence of divergent transitions
and Geweke’s convergence diagnostic~\citep{geweke1991} on
$\omega(t_k)$, $\sigma_{x_k}^{-2}$, and
$\sigma_{{\tilde{x}}_{\tilde{k}}}^{-2}$.  Sampling efficiency is
assessed with a target acceptance rate higher than .8, and the tree
depth and number of leapfrog jumps not hitting the maximum value. The
posterior expectation Monte Carlo standard error for the various
weight parameters $\omega(t_k)$, $\sigma_{x_k}^{-2}$, and
$\sigma_{{\tilde{x}}_{\tilde{k}}}^{-2}$ is monitored with a target
value of less than 10\% of the estimated posterior standard deviation.

\subsection{Validation}\label{sec:validation}

Consider a complementary collection with $H$ pairs of training and test sets,
i.e. $\{(\mathcal{D}_{h}, \mathcal{D}_{h + H}): h = 1, \dots, H\}$. For each
subset pair $h$ and plausible model $p$, we train a model on the training output
vector $\mathbf{y}$ corresponding to the training input matrix $\mathbf{X}$ and
compute the predictive mean vector and covariance matrix for the test output
vector $\mathbf{y}_*$ corresponding to the test input matrix $\mathbf{X}_*$.
Define the signal covariance matrix
$\mathbf{S}_f$ with entries ${(\mathbf{S}_f)}_{i, j} = s_f(X_i, X_j |
{\bm{\theta}})$.
Let $\predmean = \EE{\mathbf{y}_* | \mathbf{y},
  \mathbf{X}, \mathbf{X}_*}$, $\predvar = \VV{\mathbf{y}_* | \mathbf{y},
  \mathbf{X}, \mathbf{X}_*}$ be the posterior predictive mean vector and
covariance matrix computed as in
\cref{eq:predictive-mean,eq:predictive-cov} for a fixed vector of
tuned parameters $\hat{\bm{\theta}}$, where the superscripts $h,p$
were dropped for readability,
\begin{align}
  \predmean
  &= \hat{S}^{f}(\mathbf{X}_*, \mathbf{X})
    {\hat{S}^{y}(\mathbf{X}, \mathbf{X})}^{-1}
    \mathbf{y} \label{eq:predictive-mean} \\
  \predvar
  &= \hat{S}^{f}(\mathbf{X}_*, \mathbf{X}_*) 
  -
    \hat{S}^{f}(\mathbf{X}_*, \mathbf{X})
    {\hat{S}^{y}(\mathbf{X}, \mathbf{X})}^{-1}
    \hat{S}^{f}(\mathbf{X}, \mathbf{X}_*) 
  + \hat{\sigma}^2_{\varepsilon} I
    \label{eq:predictive-cov}
\end{align}

The study focuses on two validation statistics: the root mean square
error (RMSE) for the predictive mean versus the test set actual values
$(v_1)$, and the negative posterior predictive log density (negPPLD)
evaluated at the test set output values $(v_2)$. The choice of
validation statistics was motivated by the typical use of an
emulator: while some applications would only make use of the point
prediction, others combine both point prediction and predictive
uncertainty. The larger the mean square error and the predictive
variance, the more the negPPLD penalizes a model. We report the
negative log density to obtain a loss, instead of an utility function,
so that lower values of $v_1$ and $v_2$ are associated with better
prediction.  The validation statistic posterior expectation $\hat{v}$
is approximated as in \cref{eq:validation-rmse,eq:validation-ppld}
using a thinned posterior parameter sample
$\left\{\bm{\theta}_{\tilde{m}}\right\}_{\tilde{m}=1}^{\tilde{M}}$
of size $\tilde{M} << M$.
In these equations, the $h, p$ superscripts are omitted for
clarity. We compute the mean across the subset means and its
corresponding standard error, i.e.,
$\bar{v}^{(p)}
= H^{-1} \sum_{h = 1}^{H} \hat{v}^{(h, p)}$,
$\textsc{SE}\left(\bar{v}^{(p)}\right) =
{H}^{-\frac{1}{2}}
\sqrt{
  {(H - 1)}^{-1} \sum_{h = 1}^{H}
  {\left(\hat{v}^{(h, p)} - \bar{v}^{(p)}\right)}^2
}$ for every $p$.
\begingroup
\allowdisplaybreaks%
\begin{align}
  v_1
    &=
      \E_\theta\left[
      N^{-\frac{1}{2}} \norm{%
      \EE{\bm{y}_{*} | \bm{X}, \bm{X}_{*}, \bm{y}, \bm{\theta}} -
      \bm{y}_{*}
      }\right]
  = \int_{\bm{\Theta}} N^{-\frac{1}{2}} \norm{%
    \EE{\bm{y}_{*} | \bm{X}, \bm{X}_{*}, \bm{y}, \bm{\theta}} -
    \bm{y}_{*}
    }
    p(\bm{\theta} | \bm{y})
    \,\mathrm{d}\bm{\theta} \nonumber \\
  \hat{v}_1
  &= {\tilde{M}}^{-1} \sum_{{\tilde{m}}=1}^{{\tilde{M}}} N^{-\frac{1}{2}} \norm{%
    \EE{\bm{y}_{*} | \bm{X}, \bm{X}_{*}, \bm{y}, {\bm{\theta}}_{{\tilde{m}}}} -
    \bm{y}_{*}
    } \label{eq:validation-rmse} \\
  v_2
  &= p(\bm{y}_{*} | \bm{X}, \bm{X}_{*}, \bm{y}) \nonumber 
  = \int_{\bm{\Theta}}
    p(\bm{y}_{*} | \bm{X}, \bm{X}_{*}, \bm{y}, \bm{\theta}) \
    p(\bm{\theta} | \bm{y})
    \,\mathrm{d}\bm{\theta} \nonumber \\
  \hat{v}_2
  &= {\tilde{M}}^{-1} \sum_{{\tilde{m}}=1}^{{\tilde{M}}}
    p(\bm{y}_{*} | \bm{X}, \bm{X}_{*}, \bm{y},
    {\bm{\theta}}_{{\tilde{m}}}) \label{eq:validation-ppld}
\end{align}
\endgroup

A set of secondary validation statistics are reported
in the supplementary material~\cref{app:validation}, namely the negative continuous ranked
probability score (negCRPS)~\citep{gneiting2007}, the proportion of
actual values within the point-wise predictive interval, and the
coefficient of determination between actual values and the predictive
mean. The coefficient of determination is approximately equal to $1 -
v_1^2$ given that the test output variance is approximately 1. The
CRPS is within an additive constant of the PPLD under the multivariate
normality assumption. We find the CRPS more appropriate than the
Mahalanobis distance whenever the statistical model does not include
all the inputs considered by the physical model even if we expect a
relatively high signal-to-noise~\citep{bastos2009}. Although
scientists might find interest in the point-wise predictive interval
nominal coverage, the statistic misses the correlation among output
values, a feature that typically proves highly relevant for
uncertainty quantification.

\subsection{Dynamic relevance screening}\label{sec:feature-importance}

\looseness=-1
Smooth dynamic relevance, i.e., a structure where input relevance
varies smoothly over the index space, is a key assumption underlying
the \textsc{ALF} models. Naturally, it is valuable to complement the
methodology with a screening analysis to assess whether this is
reasonably consistent with the data under complete absence of prior
and model information. Additionally, a screening technique may aid in
discovering from data the characteristics needed for a suitable choice
of $\omega(t)$.
Starting off with permutation feature importance (\textsc{PFI}), a
procedure originally proposed for tree and ensemble
models~\citep{breiman2001} and later extended to a wider
family~\citep{fisher2019}, we explore how the validation statistics
react to corruptions in the input profile over blocks of the index
space. Although \textsc{PFI} is biased for correlated input
variables~\citep{hooker2021}, a likely situation given the dependence
between measurements within a functional profile, studies addressing
this have so far considered non-functional features
~\citep{strobl2007,strobl2008,nicodemus2010,hooker2021}.

To study the sensitivity of predictive accuracy to the information
contained in different index subspaces, we quantify the change in the
validation statistics as we permute pieces of the input profile. This
renders uninformative one index subspace at a time while keeping the
marginal distribution of the input unchanged. We mitigate the bias by
blocking the input values by the index, i.e., by treating all profile
measurements within an index interval as a block. Since \PFI{} was
originally conceived to compare across features, we call this
formulation permutation feature dynamic importance (\PFDI{}) as we
zoom within a profile to gauge feature importance of a single
input variable over the index space. Let
$\left\{\mathrm{T}_u\right\}_{u=1}^{U}, U \in \mathbb{N}$ form a
partition of the index space $\mathcal{T}$. Let $\mathbf{x}_{i,u} =
\{x_{i,t_k}: t_k \in \mathrm{T}_u \}$ be the input vector projected on
$\mathrm{T}_u$ for the $i$-th observation,
$\underline{\mathbf{x}}_{i,u} = [\mathbf{x}_{i,1}\cdots
\mathbf{x}_{i',u}\cdots\mathbf{x}_{i,U}]$ the corrupted input vector
where $i' \ne i$ is chosen by random permutation, and
$\underline{\mathbf{X}}_u$ the corrupted input matrix.
A model with $d_{\textsc{ARD}}(\cdot)$ and $\sigma_{x_k}^2 > 0 \ \forall \ k
= 1, \dots K$ trained on $\mathbf{X}$ is now validated on
$\underline{\mathbf{X}}_{*u}$ instead of
$\mathbf{X}_{*}$, i.e.,
$\underline{v}_{1,u} = N^{-\frac{1}{2}} \norm{%
    \EE{\bm{y}_{*} | \bm{X}, \underline{\mathbf{X}}_{*u}, \bm{y}} -
    \bm{y}_{*}
  }$ and
  $\underline{v}_{2,u} = p(\bm{y}_{*} | \bm{X},
  \underline{\bm{X}}_{*u}, \bm{y})$.
The total number of possible permutations is equal to $n!$ minus the
unique combinations of the rows observed in the original sample. The
validation statistic could be estimated as a mean over all or some
possibilities~\citep{fisher2019}, including the specific case of a
single random permutation~\citep{breiman2001}. The validation statistic
produced with the corrupted test input matrix is compared to the
reference point via the deterioration statistic $\Delta_{u} =
\underline{v}_{u} - v$.

The higher the increase in the loss statistic due to the permutation
in the input block associated with the $u$-th index interval, the more
reliant prediction accuracy is on the corresponding index
subspace. While larger $\sigma_{x_k}^{-2}$ values indicate higher
in-sample relevance at $t_k$, larger $\Delta_{u}$ empirically signals
a higher contribution of $\mathrm{T}_u$ on prediction accuracy.
When computing the functional norm in \cref{eq:functional-norm}, a
weight function $\omega(t)$ matching the \PFDI{} profile
${\left\{\Delta_{u}\right\}}_1^U$ will assign higher relevance to
input differences in those index subspaces to which model accuracy is more
sensitive. \PFDI{} may thus be used as an exploratory tool to gain
insight in the characteristics of a suitable weight function parametric
form as well as a diagnostic tool to gauge how well the trained model
reproduces the out-of-sample pattern as will be illustrated in
\cref{sec:application}.

The construction of the index partition might be motivated by the
problem or the data. In some applications, guidance can be obtained
from the physical elements involved. For example, soil nutrition
profiles can be deconstructed in layers, spectral frequencies in
bands, and time in cycles or intervals. Otherwise, data-driven
techniques may be used to identify, either manually or via algorithms
such as a hierarchical clustering, a blocking structure so that the
input correlation matrix approximates a block matrix with high
absolute correlations in the main-diagonal blocks and small absolute
correlations in the off-diagonal blocks.

\subsection{Implementation}\label{sec:implementation}

The sampler is written in the Stan probabilistic programming
language~\citep{Stan221}. The inference and analysis procedures were
implemented in the R programming language~\citep{Rcore2021} leveraging
the \texttt{rstan} interface~\citep{Rrstan2212} and other
statistical~\citep{Rmcmcse2021,Rmvtnorm2021,Rfda2021},
infrastructure~\citep{Rdatatable2021,Rdocopt2020,Rtinytest2020,Rcore2021,%
Rlogging2019,Rpbapply2021}, and
reporting~\citep{Rggplot2,RgridExtra2017,RGGally2021,Rggh4x2021,Rxtable2019}
community packages.

\section{Case study}\label{sec:application}

In this section, we illustrate how ALF weights can be used with atmospheric
profiles as functional inputs for atmospheric radiative transfer forward models.
%
%
%
%
The forward model for a clear-sky atmosphere emitting unpolarized
radiation~\citep{read2006,schwartz2006},
a two-dimensional extension of a previous model~\citep{waters1999},
%
consists of an atmospheric radiative transfer model that
calculates the radiative transfer of electromagnetic radiation through
the atmosphere.
At its core lies the unpolarized radiative transfer
equation for a nonscattering atmosphere in local thermodynamic
equilibrium, a first order partial differential equation handling the
dynamics of monochromatic, single-ray limb radiance over the
spectral frequency accounting for atmospheric absorption, and a
radiance source.
It has been implemented mainly in two computer codes, written in
FORTRAN-90 and Interactive Data Language, plus a third simplified code
used for quality assurance and verification.


Forward models are used in satellite remote sensing applications to estimate, or
\emph{retrieve}, geophysical variables from satellite observations of
electromagnetic radiation.
While methods vary across missions and applications (e.g., least squares,
regularized maximum likelihood estimation, Bayesian inference), solving this
inverse problem typically relies on iterative evaluations of the computationally
expensive code.
Similarly, studies of uncertainty through fully Bayesian retrievals and
large-scale Monte Carlo simulation experiments quickly become computationally
burdensome due to reliance on expensive forward model
evaluations~\citep{brynjarsdottir2018,lamminpaa2019,turmon2019, braverman2021}.
Naturally, attempts have been made to create an approximate, relatively faster
representation of the computer model either through surrogate models that
simplify physical assumptions while preserving some key physical
laws~\citep{hobbs2017} or via Gaussian process models treating the atmospheric
states as vector-valued inputs~\citep{johnson2020,ma2021}.

\begin{figure}
  \centering
    \includegraphics[width=1\textwidth]{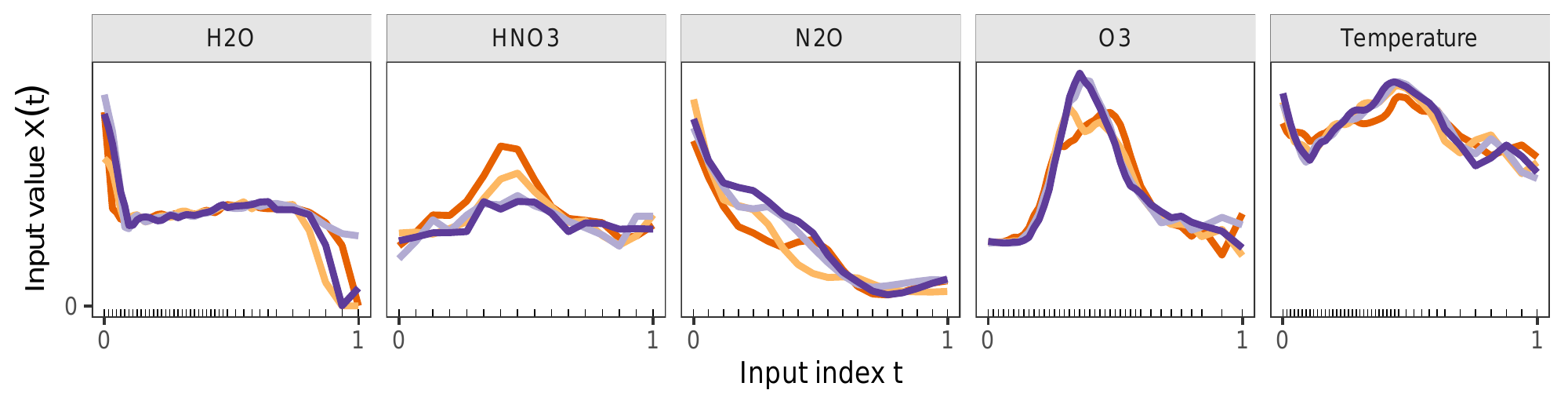}
    \caption[Four sounding input profiles]{%
      Four sounding input profiles
      $t_k\in[0,1],x_{t_k}\in[0, 1]$ for
      $k = 1, \dots, K^{(q)} \le 43, q = 1, \dots, Q = 5$.
      Although plotted as lines for ease of access, we
      observe a finite number of measurements at the locations
      indicated in the rug.}%
  \label{fig:input-profile-plot}
\end{figure}

For this case study, we consider the NASA's Microwave Limb Sounder (MLS)
mission~\citep{waters2006}. Since its launch aboard EOS-Aura in July 2004, MLS
has been producing thousands of measurements daily at fine spatial resolution on
the chemistry and dynamics of the upper troposphere, stratosphere, and
mesosphere~\citep{livesey2006,liversey2020}.
As the satellite orbits Earth, the instrument performs continual vertical scans
in the forward limb measuring thermal microwave emission in several spectral
regions containing characteristic information about temperature,
atmospheric pressure, and numerous chemicals of interest (e.g., O$_2$, O$_3$,
H$_2$O, ClO, HNO$_3$, N$_2$O, CO, OH, SO$_2$, BrO, HOCl, HO$_2$, HCN, and
CH$_3$CN).
The MLS radiative transfer forward model takes as inputs vectors containing the
aforementioned species composition at discretized vertical levels of the
atmosphere (\cref{fig:input-profile-plot}) and produces simulated radiances with
both spectral and vertical dimension in one or several vertical scans. Although
the computer code inputs and outputs are both functional, as a proof of concept
of our methodology we reduce the scope of this problem to
%
%
a scalar output and focus instead on the functional treatment of the atmospheric
input profiles.

%
%
%
%
The computer model
input corresponds to a sub-space of the complete retrieved state space
${X_{\textsc{OG}}(t)}^{(q)} \in \mathbb{R}$, where $t$ is defined
below, for $q = 1, \dots, Q = 5$ characterizing the vertical profiles
for H$_2$O, O$_3$, N$_2$O, HNO$_3$, and temperature respectively.
Out of all the retrieved species, these five are believed to be most relevant to
radiances in the electromagnetic spectrum region covered by Band 2, namely near
190 GHz~\citep{waters2006}.
For each input $q$, the state vector
$\smash{\mathbf{x}_{\textsc{OG}}^{(q)} \in \mathbb{R}^{K^{(q)}}}$ is
defined on a fixed atmospheric pressure grid
$\smash{\mathbf{t}_{\textsc{OG}}^{(q)} \in {\RealsP}^{K^{(q)}}}$
measured in hectopascals (hPa).
We restrict the analysis to pressure regions for each species that are expected
to be well-informed by the measurements as suggested
in~\citet{liversey2020}, and we thus obtain $K^{(q)} = 42, 39, 18, 16$, and $43$
respectively.
The index and the input values are normalized to the
unit interval via $g(z): \Reals \to [0, 1] = (z - l) / (u - l)$ using
the boundary values $l < u \in \Reals$ indicated in
the supplementary material \cref{tab:input-scales}, i.e., $\smash{{ {X}(t)}^{(q)} =
g({X_{\textsc{OG}}(t)}^{(q)} )}$, $\smash{\mathbf{x}^{(q)} =
g(\mathbf{x}_{\textsc{OG}}^{(q)})}$, and $\smash{\mathbf{t}^{(q)} =
g(-\log_{10} \mathbf{t}_{\textsc{OG}}^{(q)})}$.  Note that $t \to
0^{+}$ and $t \to 1^{-}$ as measurements near the tropopause and
mesopause, respectively, with smaller values indicating measurements
closer to ground. For a given species, all the observations are
evaluated at the same index, i.e., $\mathbf{t}^{(q)}$ varies per $q$
but not over the soundings $n = 1, \dots,
N$. Four sounding input profiles are illustrated in
\cref{fig:input-profile-plot}.
We define the computer model output $y = 6.82^{-1} \left(y^{\textsc{PC1}} -
.55\right) \in \mathbb{R}$ with $\EE{y}\approx0$ and $\VV{y}\approx1$, where
$y^{\textsc{PC1}}\in\mathbb{R}$ denote the MLS Band 2 radiances first
multivariate functional principal component score~\citep{happ2018}, ordered by
largest eigenvalue, as produced in related work by~\citet{johnson2020}. All the
scaling constants were obtained empirically. The data set is given by the
collection $\mathcal{D} = {\left\{ y_n, t_{k^{(q)}}^{(q)}, x_{n, k^{(q)}}^{(q)}
\right\}}_{n, q, k^{(q)}}$.
%


A total of $16,000$ samples are randomly partitioned into $H = 8$
training and $H$ test complementary subsets with $N = 1,000$ soundings
each.  We evaluate the performance of all three \textsc{ALF}
models described in \cref{sec:estimation}. For comparison, we also
consider the vector-input Gaussian process (\viGP) models
\textsc{viGP-SE} with $d_{\textsc{ARD}}(\cdot)$ and $\sigma_{x_k}^2
= \sigma_{x}^2 > 0 \ \forall \ k = 1, \dots K$;
\textsc{viGP-ARD} with $d_{\textsc{ARD}}(\cdot)$ and $\sigma_{x_k}^2
> 0 \ \forall \ k = 1, \dots K$;
\textsc{viGP-FPCA} with $d_{\textsc{PC}}(\cdot)$ and
$\sigma_{{\tilde{x}}_{\tilde{k}}}^2 > 0 \ \forall \ \tilde{k} = 1,
\dots \tilde{K}_{\textsc{99\%}}$;
\textsc{viGP-FFPCA} with $d_{\textsc{PC}}(\cdot)$ and
$\sigma_{{\tilde{x}}_{\tilde{k}}}^2 > 0 \ \forall \ \tilde{k} = 1,
\dots, \tilde{K}_{\textsc{full}}$.
Models \textsc{viGP-SE} and \textsc{viGP-ARD} have a shared and a separable
correlation structure on the vector space respectively. Models
\textsc{viGP-FPCA} and \textsc{viGP-FFPCA} have a separable correlation
structure on the reduced and full principal component space respectively, the
former being restricted to the components accounting for 99\% of the input
variability. FPCA is set up with a cubic spline smoother, 10 equally-spaced
knots in [0, 1], and 12 basis functions, thus $\tilde{K}_{\textsc{full}} = 12$.
One model is fit separately for each atmospheric input.

Specifying the prior families in \cref{sec:estimation}, we set the
following independent priors:
$\phi \sim \textsc{InvGamma}(5, 5)$,
$\tau \sim \textsc{Beta}(1, 1)$,
$\smash{{10}^{-\frac{1}{2}} \lambda \sim \textsc{N}^{+}(0, 1)}$,
$\log(\kappa) \sim \textsc{N}(0, 1)$ for the \fiGP{} parameters,
$\sigma_{x},
\sigma_{x_k}, \sigma_{{\tilde{x}}_{\tilde{k}}} \iid
\textsc{InvGamma}(5, 5)$ for the \viGP{} parameters, and
$\sigma_f, \sigma_{\varepsilon} \iid \textsc{N}^{+}(0, 1)$ for the
signal and noise standard deviation parameters. The log posterior is evaluated at
3,000 random locations of the parameter space drawn from $\sigma_{x},
\sigma_{x_k}, \sigma_{{\tilde{x}}_{\tilde{k}}}, \phi, 5^{-1}
\sigma_f, 2\sigma_{\varepsilon} \iid \textsc{N}^{+}(0, 1)$,
$\lambda_1, \lambda_2 \iid \textsc{C}^{+}(0, 1)$, and $\tau \sim
\textsc{Beta}(2/3, 1)$. A beta distribution with mean equal to $0.4$
reflects the intuition that input values near ground are expected to
have relatively higher predictive relevance.
The standard deviations are compatible with the output scale.
The multiplicative factors for the standard deviations are
motivated by the signal-to-noise ratio $\textsc{STN} = \sigma_f /
\sigma_{\varepsilon}$ observed in previous exploratory studies. A
total of 30 optimizations are initialized at the highest posterior
density parameter vectors selected from the 3,000 random candidates.
Finally, the optimal values with highest posterior density are used to
initialize one MCMC chain with 500 warm-up iterations and $M = 1,500$ post
warm-up samples. Validation statistics are computed using a sample of $\tilde{M}
= 100$ thinned observations sampled with a batch size of 150 iterations.

\begin{figure}
  \centering
  \includegraphics[width=.69\textwidth]{%
    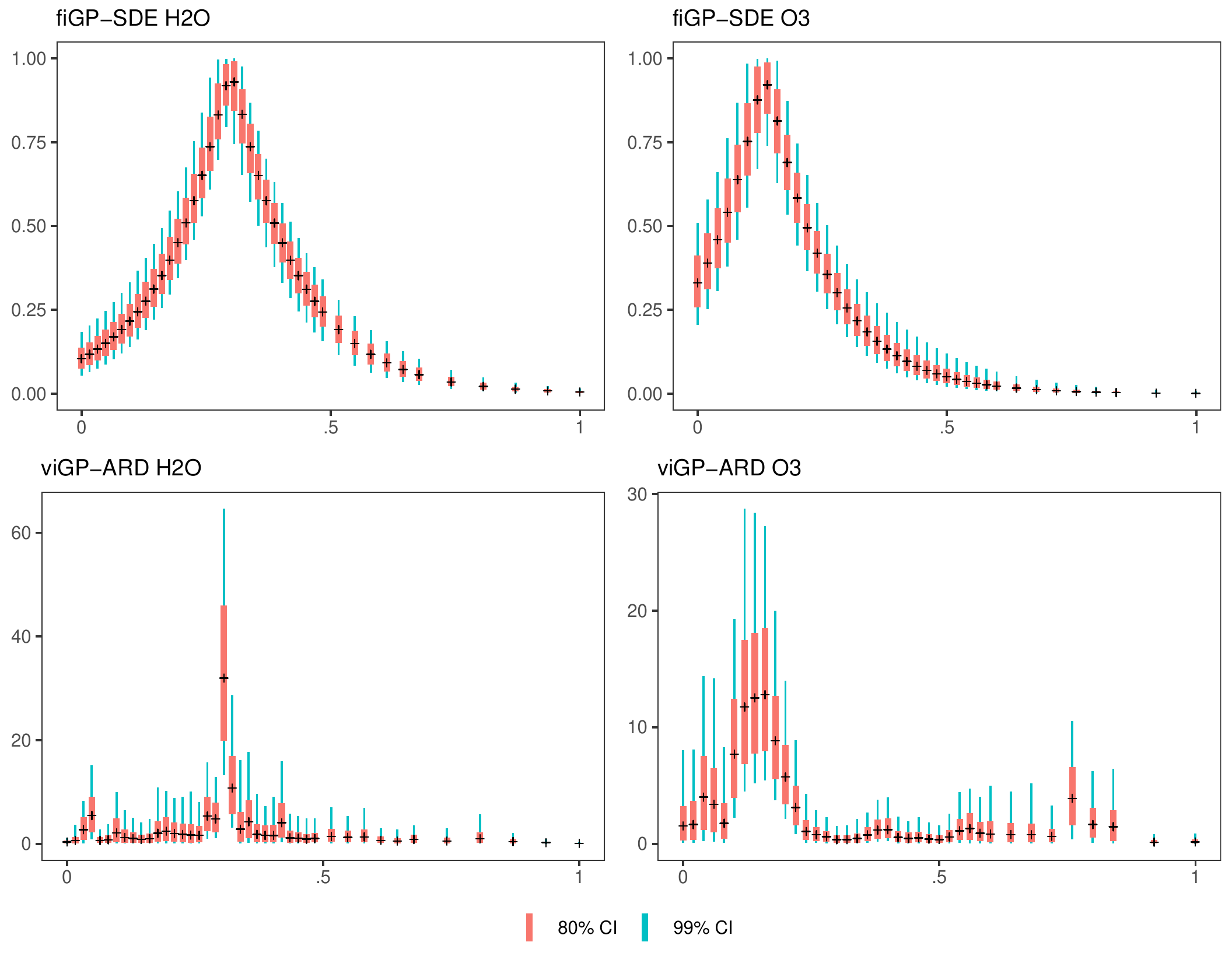}
  \caption[Model weights]{Weight posterior intervals for two models
    and inputs.}%
  \label{fig:alf-weight-posteriors}
\end{figure}
We now discuss the general patterns in the posterior density of the
parameter vector $\bm{\theta}$ and the weight function evaluations
$\{\omega(t_k)\}_{k}$, and we then compare the prediction performance
across the plausible models. Looking at the \textsc{ADE} parameter
posterior expectation across all data subsets in
\cref{tab:validation-statistics-full}, a global pattern emerges where
the peak relevance $\tau$ is situated in the first half of the index
interval. The models posit slowly decaying rates at the left of the
peak and an order-of-magnitude faster speed toward the right, i.e.,
$\lambda_1 << \lambda_2$.  In other words, relevance is higher and
relatively more constant near ground while it decreases fast as
altitude increases above $\tau$.
Consider, for example,
\cref{fig:weight-posterior-all}.
With roughly
$\EE{\omega(t) | \mathbf{y}} < .05 \ \forall \ t > 0.5$, the posterior density
suggests that differences in H$_2$O and O$_3$ near the mesopause have
negligible predictive relevance.
The parameter posterior density for these are reported in the supplementary
material \cref{app:param-posterior}.
It might be tempting to interpret
the relevance profiles in terms of predictive power~\citep{neal1996}
and conclude that the chemical composition near the atmospheric edge
does not contribute to radiance prediction. However, it is worth
cautioning that non-linear terms are also associated with shorter
length-scales~\citep{piironen2016}, and thus the tuned parameters could
be partly capturing the fact that the relationship between the
radiance and the input is non-linear (``more wiggly'') at lower
altitudes and becomes more linear as we ascend over the vertical
profile.  The standard deviation posterior intervals confirm high
signal-to-noise ratios even when the models have only one input
variable, with noise contributing less than .05 of the total variance
for these two inputs.

\spacingset{1}
\begin{landscape}

  \begin{figure}[H]
    \centering
    \includegraphics[width=\linewidth]{%
      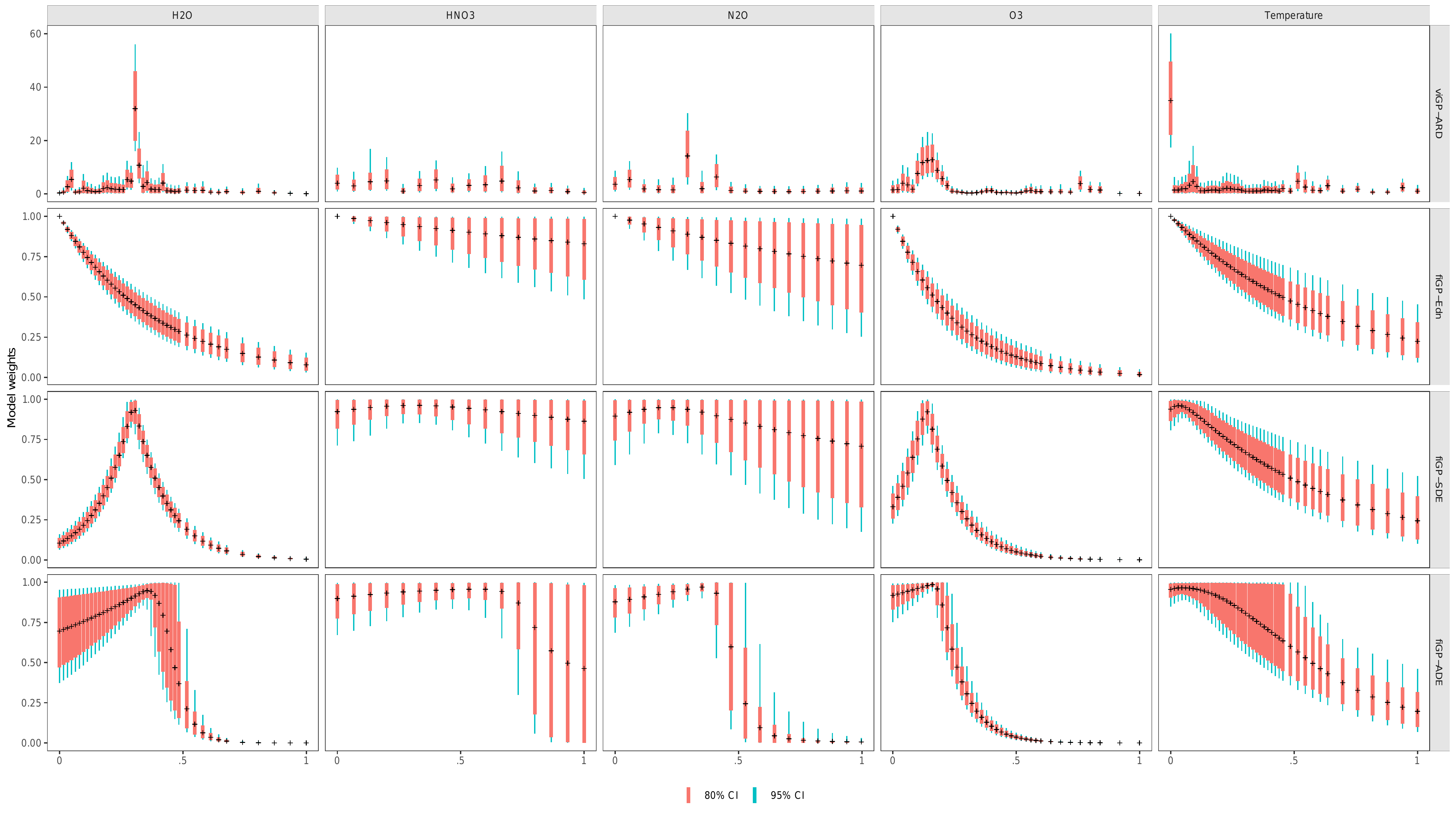}
    \caption[Model weights]{Weight posterior intervals estimated from
      one data subset. Each model (facet row) is trained to predict
      radiance as a function of each input variable (facet column)}%
    \label{fig:weight-posterior-all}
  \end{figure}



  \begin{table}[H]
    \begin{adjustbox}{width=.95\linewidth,center}
      \centering
      \begin{tabular}{lrrrrrrrrrrrrrr}
        \toprule
Model & $\tau$ & $\lambda_1$ & $\lambda_2$ & $\theta$ & $\sigma_{\varepsilon}$ & $\sigma_{f}$ & $\sigma_{f} / \sigma_{\varepsilon}$ & $p(\mathbf{\theta} | \mathbf{y})$ & $R^2$ & RMSE & Neg. CRPS & Neg. PPLD & Cov. 95\% \\ 
  \midrule
\multicolumn{14}{l}{\textbf{H2O}}\\
viGP-SE & - & - & - & - & .30 (.003) &  4.55 (.176) & 15.3 (.76) & 462 ( 2.3) & .89 (.003) & .34 (.003) & .19 (.001) & 273 ( 7.0) & .96 (.002) \\ 
  viGP-ARD & - & - & - & - & .28 (.003) &  5.07 (.103) & 17.8 (.52) & 320 ( 5.4) & \textbf{.91} (.003) & \textbf{.31} (.003) & \textbf{.17} (.001) & \textbf{196} ( 6.9) & .96 (.003) \\ 
  viGP-FPCA & - & - & - & - & .66 (.009) & .63 (.013) &  1.0 (.02) & -122 ( 9.7) & .55 (.010) & .67 (.010) & .37 (.006) & 1024 (14.1) & .93 (.003) \\ 
  viGP-FFPCA & - & - & - & - & .36 (.004) & .82 (.010) &  2.3 (.03) & 256 ( 6.6) & .79 (.014) & .46 (.015) & .25 (.007) & 535 ( 6.8) & .93 (.006) \\ 
  fiGP-Edn & - & - &  2.40 (.071) & .09 (.002) & .31 (.003) &  6.46 (.152) & 21.1 (.73) & 419 ( 2.1) & .89 (.003) & .33 (.004) & .18 (.001) & 261 ( 7.5) & .96 (.002) \\ 
  fiGP-SDE & .30 (.001) &  7.59 (.193) &  7.59 (.193) & .06 (.001) & .29 (.003) &  6.70 (.096) & 23.0 (.51) & \textbf{470} ( 4.9) & \textbf{.91} (.003) & \textbf{.31} (.004) & \textbf{.17} (.001) & \textbf{202} ( 7.1) & .96 (.002) \\ 
  fiGP-ADE & .38 (.015) &  1.69 (.359) & 15.70 (.793) & .07 (.002) & .29 (.003) &  6.60 (.120) & 22.7 (.66) & \textbf{479} ( 5.0) & \textbf{.90} (.003) & \textbf{.31} (.004) & \textbf{.17} (.001) & \textbf{202} ( 6.5) & .96 (.003) \\ 
   \midrule
\multicolumn{14}{l}{\textbf{HNO3}}\\
viGP-SE & - & - & - & - & .41 (.011) &  1.30 (.051) &  3.2 (.11) & \textbf{231} (16.7) & \textbf{.77} (.004) & \textbf{.48} (.005) & \textbf{.26} (.003) & \textbf{614} (10.4) & .95 (.002) \\ 
  viGP-ARD & - & - & - & - & .43 (.007) &  2.77 (.068) &  6.5 (.15) & 140 (14.3) & \textbf{.78} (.004) & \textbf{.47} (.005) & \textbf{.25} (.003) & \textbf{619} (11.6) & .94 (.003) \\ 
  viGP-FPCA & - & - & - & - & .90 (.002) & .87 (.034) &  1.0 (.04) & -416 ( 2.3) & .19 (.006) & .91 (.007) & .51 (.004) & 1320 ( 7.7) & .95 (.004) \\ 
  viGP-FFPCA & - & - & - & - & .35 (.009) & .94 (.014) &  2.7 (.08) & 150 (14.2) & .71 (.007) & .54 (.007) & .28 (.003) & \textbf{646} ( 7.7) & .95 (.003) \\ 
  fiGP-Edn & - & - & .42 (.080) & .12 (.005) & .43 (.009) &  2.35 (.095) &  5.4 (.14) & \textbf{199} (15.1) & \textbf{.78} (.004) & \textbf{.47} (.005) & \textbf{.25} (.003) & \textbf{623} (10.9) & .95 (.003) \\ 
  fiGP-SDE & .29 (.025) & .69 (.175) & .69 (.175) & .12 (.005) & .43 (.009) &  2.39 (.105) &  5.5 (.17) & \textbf{198} (15.0) & \textbf{.78} (.004) & \textbf{.47} (.005) & \textbf{.25} (.003) & \textbf{623} (10.9) & .95 (.003) \\ 
  fiGP-ADE & .62 (.013) & .22 (.017) & 24.13 ( 3.011) & .12 (.006) & .43 (.007) &  2.58 (.116) &  6.0 (.21) & \textbf{204} (14.2) & \textbf{.78} (.003) & \textbf{.47} (.005) & \textbf{.25} (.003) & \textbf{610} (11.2) & .95 (.003) \\ 
   \midrule
\multicolumn{14}{l}{\textbf{N2O}}\\
viGP-SE & - & - & - & - & .42 (.004) &  1.04 (.043) &  2.5 (.10) & \textbf{291} ( 9.8) & \textbf{.81} (.004) & \textbf{.44} (.005) & \textbf{.24} (.002) & \textbf{585} (10.3) & .94 (.002) \\ 
  viGP-ARD & - & - & - & - & .42 (.004) &  1.93 (.074) &  4.6 (.17) & 194 ( 9.8) & \textbf{.81} (.004) & \textbf{.43} (.005) & \textbf{.24} (.002) & \textbf{581} (11.0) & .94 (.001) \\ 
  viGP-FPCA & - & - & - & - & .97 (.005) & .53 (.020) & .5 (.02) & -484 ( 5.1) & .04 (.006) & .99 (.008) & .56 (.004) & 1406 ( 7.9) & .94 (.003) \\ 
  viGP-FFPCA & - & - & - & - & .41 (.004) & .64 (.008) &  1.6 (.02) & 160 ( 9.1) & \textbf{.79} (.008) & \textbf{.46} (.009) & .25 (.004) & \textbf{630} (12.9) & .94 (.003) \\ 
  fiGP-Edn & - & - & .66 (.066) & .14 (.002) & .43 (.004) &  2.30 (.074) &  5.4 (.18) & \textbf{261} ( 9.6) & \textbf{.81} (.004) & \textbf{.44} (.005) & \textbf{.24} (.002) & \textbf{585} (10.6) & .94 (.001) \\ 
  fiGP-SDE & .24 (.018) & .75 (.063) & .75 (.063) & .14 (.003) & .43 (.004) &  2.26 (.073) &  5.3 (.19) & \textbf{259} ( 9.6) & \textbf{.81} (.004) & \textbf{.44} (.005) & \textbf{.24} (.002) & \textbf{585} (10.6) & .94 (.001) \\ 
  fiGP-ADE & .43 (.005) & .35 (.043) & 28.19 ( 2.163) & .13 (.003) & .43 (.004) &  2.53 (.061) &  5.9 (.16) & \textbf{266} ( 9.7) & \textbf{.81} (.004) & \textbf{.43} (.005) & \textbf{.24} (.002) & \textbf{581} (11.0) & .94 (.001) \\ 
   \midrule
\multicolumn{14}{l}{\textbf{O3}}\\
viGP-SE & - & - & - & - & .24 (.004) &  6.02 (.244) & 25.4 ( 1.11) & \textbf{540} ( 7.4) & .90 (.004) & .32 (.007) & .17 (.003) & 138 (10.1) & .97 (.002) \\ 
  viGP-ARD & - & - & - & - & .24 (.003) &  6.85 (.140) & 28.8 (.73) & 387 ( 9.9) & \textbf{.91} (.002) & \textbf{.30} (.003) & \textbf{.16} (.001) & \textbf{92} ( 8.7) & .96 (.001) \\ 
  viGP-FPCA & - & - & - & - & .45 (.004) &  1.19 (.052) &  2.7 (.12) & 260 (10.9) & .79 (.003) & .46 (.006) & .25 (.002) & 637 (13.0) & .93 (.002) \\ 
  viGP-FFPCA & - & - & - & - & .24 (.004) & .83 (.014) &  3.5 (.09) & 489 ( 9.9) & .86 (.004) & .38 (.006) & .20 (.003) & 295 (11.2) & .96 (.001) \\ 
  fiGP-Edn & - & - &  4.02 (.143) & .07 (.001) & .25 (.004) &  8.22 (.137) & 33.6 (.83) & \textbf{528} ( 8.9) & \textbf{.92} (.003) & \textbf{.29} (.004) & \textbf{.16} (.002) & \textbf{90} ( 9.0) & .96 (.002) \\ 
  fiGP-SDE & .14 (.002) &  7.17 (.390) &  7.17 (.390) & .07 (.001) & .25 (.004) &  8.21 (.154) & 33.2 (.82) & \textbf{543} (10.3) & \textbf{.92} (.002) & \textbf{.29} (.004) & \textbf{.15} (.002) & \textbf{85} ( 8.9) & .96 (.002) \\ 
  fiGP-ADE & .24 (.031) & .48 (.062) & 10.59 (.621) & .07 (.003) & .25 (.004) &  8.16 (.158) & 33.0 (.94) & \textbf{553} (10.5) & \textbf{.92} (.002) & \textbf{.29} (.004) & \textbf{.16} (.001) & \textbf{87} ( 8.7) & .96 (.003) \\ 
   \midrule
\multicolumn{14}{l}{\textbf{Temperature}}\\
viGP-SE & - & - & - & - & .21 (.003) &  1.27 (.034) &  5.9 (.15) & \textbf{798} ( 9.3) & \textbf{.94} (.001) & \textbf{.25} (.002) & \textbf{.14} (.001) & \textbf{-7} ( 1.6) & .97 (.003) \\ 
  viGP-ARD & - & - & - & - & .23 (.003) &  2.82 (.030) & 12.5 (.20) & 559 ( 8.4) & \textbf{.94} (.001) & \textbf{.25} (.002) & \textbf{.14} (.001) & \textbf{-13} ( 3.7) & .96 (.002) \\ 
  viGP-FPCA & - & - & - & - & .53 (.009) &  1.07 (.031) &  2.0 (.09) &  85 (16.6) & .71 (.008) & .54 (.008) & .29 (.003) & 802 (14.5) & .94 (.004) \\ 
  viGP-FFPCA & - & - & - & - & .28 (.005) & .76 (.013) &  2.7 (.09) & 467 (10.2) & .89 (.001) & .33 (.002) & .18 (.001) & 268 ( 7.2) & .96 (.003) \\ 
  fiGP-Edn & - & - &  1.31 (.129) & .08 (.003) & .23 (.003) &  2.94 (.094) & 12.8 (.39) & 729 ( 8.9) & \textbf{.94} (.001) & \textbf{.25} (.003) & \textbf{.14} (.001) & \textbf{4} ( 5.3) & .96 (.002) \\ 
  fiGP-SDE & .08 (.015) &  1.39 (.151) &  1.39 (.151) & .08 (.003) & .23 (.003) &  2.92 (.106) & 12.8 (.43) & 726 ( 8.9) & \textbf{.94} (.001) & \textbf{.25} (.003) & \textbf{.14} (.001) & \textbf{4} ( 5.4) & .96 (.003) \\ 
  fiGP-ADE & .34 (.030) & .35 (.096) &  3.09 (.429) & .08 (.003) & .23 (.003) &  2.86 (.096) & 12.6 (.37) & 729 ( 9.1) & \textbf{.94} (.002) & \textbf{.25} (.003) & \textbf{.14} (.002) & \textbf{2} ( 5.2) & .96 (.003) \\ 
   \bottomrule
\multicolumn{14}{l}{}\\

      \end{tabular}
    \end{adjustbox}
    \caption{
      Tuned model parameters and validation statistics
      across five input variables (line-separated table sections) and
      plausible models (rows).
      Each cell reports the mean across the subset means $\bar{v}^{(p,q)}$
      and the corresponding standard error $\textsc{SE}\left(\bar{v}^{(p,
          q)}\right)$ in parenthesis. Bold values are the best in
      class.%
    }\label{tab:validation-statistics-full}
  \end{table}

  \clearpage
\end{landscape}
\spacingset{2}

It is also of interest to compare the posterior intervals for the
\textsc{SDE} and \textsc{ARD} weights, i.e., $\omega(t_k)$ versus
$\sigma_{x_k}^{-2}$ over $k$. The latter offers full flexibility at
the price of $K^{(q)} / 3 < 14$ times more tuning parameters. In
\cref{fig:alf-weight-posteriors}, we observe that
%
both the parametric and the fully separable
weights estimate the highest relevance index region in the same
neighborhood.  For H$_2$O, $\argmax_k \EE{\sigma_{x_k}^{-2} | \mathbf{y}}
= 20$ with $t_{20} = .31$, and $\EE{\tau | \mathbf{y}} = .30$ with
$P(.27 \le \tau \le .32 | \mathbf{y}) = .95$. Similarly for O$_3$,
$\argmax_k \EE{\sigma_{x_k}^{-2} | \mathbf{y}} = 9$ with $t_{9} =
.16$, and $\EE{\tau | \mathbf{y}} = .13$ with $P(.11 \le \tau \le .16
| \mathbf{y}) = .95$. The \textsc{ARD} weights are locally higher near
the maximum and vary rather smoothly across $k$, just as assumed by
\textsc{ALF}.
However, posterior uncertainty in the relevance profiles manifest
rather homogeneously over the index space for \textsc{SDE} whereas
\textsc{ARD} exhibits increased uncertainty around the relevance peak
with a coefficient of variation 5 times larger than in \textsc{SDE}
(.06 and .32 respectively).

The validation statistics
$\mathcal{V} = \left\{\bar{v}^{(p, q)}\right\}$ are summarized in
\cref{tab:validation-statistics-mini}. See
\cref{tab:validation-statistics-full,fig:validation-foldmeans-brief}
in the supplementary material for a detailed report of in and out-of-sample
statistic means and standard errors.
The across-input means
$\bar{v}^{(p)} = 5^{-1} \sum_{q = 1}^{5} \bar{v}^{(p, q)}$ suggest
that the \textsc{ALF} models perform similarly to the \textsc{viGP}
models with no input pre-processing. For most purposes, there is a
negligible edge for \textsc{SDE} and \textsc{ADE} over \textsc{Edn},
and \textsc{ARD} over \textsc{SE}. With the aid of
\cref{tab:validation-statistics-full}, we observe that this difference
is slightly more favorable for H$_2$O and Temperature.  Overall, the
\textsc{ADE} and \textsc{ARD} models are arguably a sensible choice
across this study combinations.  In most situations, it is also
reasonable to argue for the \textsc{SDE} variant given the similar
predictive quality and parsimony. The coefficient of determination and
the negative CRPS produce similar findings to the RMSE and the
negative PPLD, as expected given that these quantities are
intrinsically related. Coupling \textsc{ADE} or \textsc{ARD} with
temperature produces the highest coefficient of determination .94, the
smallest RMSE .25, and CRPS .14 and the best PPLD.~As a general note,
the actual mean coverage of the point-wise predictive interval is
close to its nominal value in most cases.  For all inputs,
\textsc{FFPCA} performs worse than \textsc{ARD} and
\textsc{fiGP}. Additionally, \textsc{FFPCA}'s RMSE is 17-54\% smaller
than \textsc{FPCA}'s, which retains 99\% of the explained
variation. This is yet another illustration that variability in the
input space need not relate with predictive power.
\begin{table}
  \adjustbox{width=.79\textwidth,center}{%
    \centering
    \begin{tabular}{lrrrrr|r}
      \toprule
 & H2O & HNO3 & N2O & O3 & Temp & Mean \\ 
  \midrule
  \textsc{SE}    &  .34       &  {\bf .48} &  {\bf .44} &  .32       &  {\bf .25} &  .37 \\ 
  \textsc{ARD}   &  {\bf .31} &  {\bf .47} &  {\bf .43} &  {\bf .30} &  {\bf .25} &  .35 \\ 
  \textsc{FPCA}  &  .67       &  .91       &  .99       &  .46       &  .54       &  .71 \\ 
  \textsc{FFPCA} &  .46       &  .54       &  {\bf .46} &  .38       &  .33       &  .44 \\ 
  \textsc{Edn}   &  .33       &  {\bf .47} &  {\bf .44} &  {\bf .29} &  {\bf .25} &  .36 \\ 
  \textsc{SDE}   &  {\bf .31} &  {\bf .47} &  {\bf .44} &  {\bf .29} &  {\bf .25} &  .35 \\ 
  \textsc{ADE}   &  {\bf .31} &  {\bf .47} &  {\bf .43} &  {\bf .29} &  {\bf .25} &  .35 \\ 
   \midrule
   Mean          &  .39 &  .55 &  .52 &  .33 &  .31 &  .42 \\ 
   \bottomrule
    \end{tabular}
    \begin{tabular}{lrrrrr|r}
      \toprule
                 & H2O & HNO3 & N2O & O3 & Temp & Mean \\ 
  \midrule
  \textsc{SE}    & 273       & {\bf 614} & {\bf 585} & 138      & {\bf -7}  & 323 \\ 
  \textsc{ARD}   & {\bf 196} & {\bf 619} & {\bf 581} & {\bf 92} & {\bf -13} & 295 \\ 
  \textsc{FPCA}  & 1024      & 1320      & 1406      & 637      & 802       & 1038 \\ 
  \textsc{FFPCA} & 535       & {\bf 646} & {\bf 630} & 295      & 268       & 475 \\ 
  \textsc{Edn}   & 261       & {\bf 623} & {\bf 585} & {\bf 90} & {\bf 4}   & 312 \\ 
  \textsc{SDE}   & {\bf 202} & {\bf 623} & {\bf 585} & {\bf 85} & {\bf 4}   & 300 \\ 
  \textsc{ADE}   & {\bf 202} & {\bf 610} & {\bf 581} & {\bf 87} & {\bf 2}   & 297 \\ 
   \midrule
   Mean & 385 & 722 & 708 & 204 & 152 & 434 \\ 
   \bottomrule
    \end{tabular}}
  \caption{Mean validation statistics
    $\bar{v}^{(p, q)}$: RMSE (left) and negPPLD (right).
    Smaller values are better. Bold is best in class.
    Note that $\EE{y}\approx 0$ and $\VV{y}\approx 1$.
  }%
  \label{tab:validation-statistics-mini}
\end{table}

\begin{figure}
  \centering
  \includegraphics[width=\textwidth]{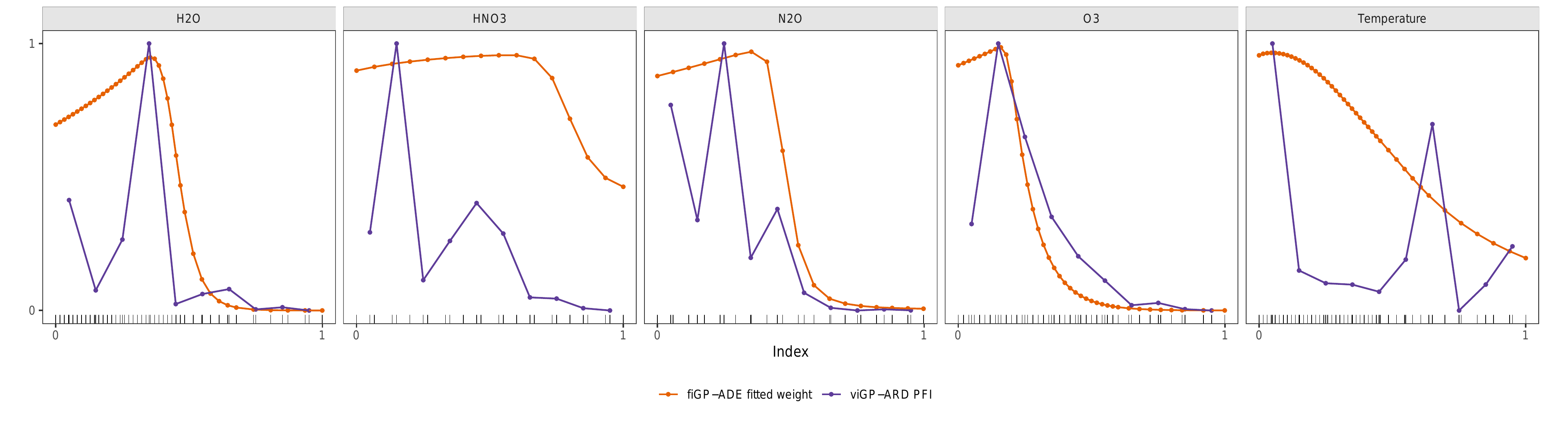}
  \caption[]{
    Weight posterior expectation $\EE{\omega(t) | \bm{X}, \bm{y}, \bm{\theta}}$
    from the \fiGPADE~model and permutation feature dynamic importance score
    normalized to one $g(\Delta_{u})$ for the \textsc{negLLPD} validation statistic
    $v_2$. Rug ticks indicate the index value at which the inputs are observed.
  }%
  \label{fig:pfi-ard-normalized}
\end{figure}

We apply the screening analysis outlined in
\cref{sec:feature-importance} and compare the weight function
posterior distribution versus the \PFDI{} coefficients. Let
$\mathrm{T}_u = \{u - 1 < 10 \ t \le u\}$ for $u = 1, \dots,
10$. Recalling that the index $t$ is the inverse vertical pressure in
$\log_{10}$ scale normalized to the unit interval, an equidistant
partition of size 10 seems sensible.  We compare the in-sample
relevance, quantified by the posterior expectation of the
\fiGPADE~weights, versus the predictive importance scores computed on
the negative posterior predictive log density $v_2$ and normalized to
1. The statistics averaged over the $H$ data sets are illustrated in
\cref{fig:pfi-ard-normalized}.  The H$_2$O and O$_3$ profiles show great
conformity. \PFDI{} reaches a maximum at $t \in (.3, .4]$ and $t \in
(.1, .2]$ while \fiGPADE~estimates the expected weight peak at $\tau =
.38$ and $\tau =.24$ respectively. Also, both methods assign higher
relative relevance to the left side of their peaks, agreeing on
atmospheric layers near ground being indeed more relevant to predict
the output scalar. HNO$_3$ and N$_2$O curves are also well aligned with
close peaks, high relative relevance toward the left and rapidly
decaying relevance as $t$ increases.
\PFDI{} uncovers multiple modes for temperature at $(0,
.1]$, $(.6, .7]$ and $(.9, 1]$ whereas  \fiGPADE{}, limited
to unimodal patterns, connects them smoothly.

The \textsc{ALF} curves are notably smoother than
\textsc{PFDI}, a desired effect to smooth out patterns of erratic
changes that might be influenced by the noise in the data rather than
the physical process itself. Overall, \PFDI{} as a diagnostic tool
suggests that \fiGPADE{} consistently learns from training data to
assign higher relevance to the intervals with highest out-of-sample
relevance while also ensuring smoother dynamics. The 
multimodality in temperature hints that future work should explore
new parametric forms for $\omega(t)$.

\section{Discussion}\label{sec:discussion}

%
To facilitate the emulation of computer experiments with functional
inputs, we proposed a new family of functional weights. The
flexibility of \textsc{ALF} weights renders it more suitable for a data-driven
approach whenever
it is reasonable to assume unimodal smoothly decaying predictive
relevance but little is known about its dynamics.
We presented a fully Bayesian inference framework via MCMC, including
a set of weakly informative priors for the \textsc{ALF} parameters.
Additionally, we adapted permutation feature importance to account for
the correlation induced by the input structure. \PFDI{} can be used as
an exploratory tool to design new parametric forms of $\omega(t)$ as
well as a diagnostic procedure to compare estimated relevance versus
out-of-sample predictive power.

We study the case of a computer experiment for the retrieval of
atmospheric parameters from thermal remote sensing data.
We observe that the \textsc{ARD} and \textsc{fiGP} models showed
similar predictive performance while functional principal
component models underperformed. Compared to a vector-input Gaussian
Process, \textsc{ALF} smooths out erratic patterns in the length-scale
parameters and offers an alternative for high definition inputs with up
to three parameter per functional input irrespective of the number of
measurements per profile without the need to reduce the input
space. \textsc{ALF} provides a simplified representation that addresses
interpretation, smoothness differentiation, and parsimony in relevance
determination.

Under the hypothesis that \textsc{ARD} might be more susceptible to
overfitting for large $K$, it would be worth exploring whether this
predictive parity holds as the resolution of the functional input
increases.  A natural extension of the \textsc{fiGP} methodology
revolves around the development of new and more flexible families of
weight functions. For example, exponential weights that are quadratic
on $t - \tau$ could avoid potential inefficiencies in some inference
algorithms due to \textsc{ALF} being non-differentiable at $t =
\tau$. Alternatives outside the exponential families includes Fourier
forms. 
Some applications might motivate multimodal relevance profiles that
could be modeled, say, as a mixture of functions or wavelet
decomposition.  Ultimate flexibility could be achieved by setting a
Gaussian process prior on the $\omega(t)$ function, or on the input
$X(t)$ itself, possibly reframing the surrogate as a
DeepGP~\citep{damianou2013,sauer2022a,sauer2022}.

Since optimizations over the log posterior surface were carried out to
initialize the MCMC chains, comparisons could be drawn between the optimal
values treated as MAP estimates and the fully Bayesian out of sample
predictions. More generally, there is an evident opportunity to couple the
\textsc{fiGP} methodology with approximate inference methods such as local
GP~\citep{gramacy2015} or the Vecchia
approximation~\citep{vecchia1988,katzfuss2021}.

\section*{Acknowledgments}\label{sec:acknowledgement}

\if0\blind{} {
  We thank the MLS team at Jet Propulsion Laboratory for their insight in the
  instrument, the forward model, and other relevant atmospheric science concepts.
  LD and JN:~Funding was partially provided by Iowa State University through the
  Presidential Interdisciplinary Research Initiative (PIR) on C-CHANGE:~Science
  for a Changing Agriculture, and the Foundation for Food and Agriculture
  Research.
  LD:~The research reported in this paper is partially supported by the HPC@ISU
  equipment at Iowa State University, some of which has been purchased through
  funding provided by NSF under MRI grant number 1726447.
  JN:~This article is a product of the Iowa Agriculture and Home Economics
  Experiment Station, Ames, Iowa. Project No. IOW03717 is supported by USDA/NIFA
  and State of Iowa funds. Any opinions, findings, conclusions, or recommendations
  expressed in this publication are those of the author(s) and do not necessarily
  reflect the views of the U.S. Department of Agriculture nor Iowa State
  University.
  MJ and JT:~Part of this research was carried out at the Jet Propulsion
  Laboratory, California Institute of Technology, under a contract with the
  National Aeronautics and Space Administration (80NM0018D0004).
} \fi

\if1\blind{} { \textrm{REDACTED} } \fi




\spacingset{1}
\bibliographystyle{apalike}
\bibliography{references,referencesR}

\begin{thebibliography}{}

\bibitem[Auguie, 2017]{RgridExtra2017}
Auguie, B. (2017).
\newblock {\em gridExtra: Miscellaneous Functions for "Grid" Graphics}.
\newblock R package version 2.3.

\bibitem[Bastos and O'Hagan, 2009]{bastos2009}
Bastos, L.~S. and O'Hagan, A. (2009).
\newblock Diagnostics for {{Gaussian}} process emulators.
\newblock {\em Technometrics}, 51(4):425--438.

\bibitem[Bayarri et~al., 2007]{bayarri2007a}
Bayarri, M.~J., Walsh, D., Berger, J.~O., Cafeo, J., {Garcia-Donato}, G., Liu,
  F., Palomo, J., Parthasarathy, R.~J., Paulo, R., and Sacks, J. (2007).
\newblock Computer model validation with functional output.
\newblock {\em The Annals of Statistics}, 35(5):1874--1906.

\bibitem[Betancourt et~al., 2020a]{betancourt2020}
Betancourt, J., Bachoc, F., Klein, T., Idier, D., Pedreros, R., and Rohmer, J.
  (2020a).
\newblock Gaussian process metamodeling of functional-input code for coastal
  flood hazard assessment.
\newblock {\em Reliability Engineering \& System Safety}, 198:106870.

\bibitem[Betancourt et~al., 2020b]{betancourt2020a}
Betancourt, J.~D., Bachoc, F., and Klein, T. (2020b).
\newblock Gaussian process regression for scalar and functional inputs with
  {{funGp}} - the in-depth tour.

\bibitem[Braverman et~al., 2021]{braverman2021}
Braverman, A., Hobbs, J., Teixeira, J., and Gunson, M. (2021).
\newblock Post hoc uncertainty quantification for remote sensing observing
  systems.
\newblock {\em SIAM/ASA Journal on Uncertainty Quantification},
  9(3):1064--1093.

\bibitem[Breiman, 2001]{breiman2001}
Breiman, L. (2001).
\newblock Random forests.
\newblock {\em Machine Learning}, 45(1):5--32.

\bibitem[Brynjarsdottir et~al., 2018]{brynjarsdottir2018}
Brynjarsdottir, J., Hobbs, J., Braverman, A., and Mandrake, L. (2018).
\newblock Optimal {{Estimation Versus MCMC}} for {{CO2 Retrievals}}.
\newblock {\em Journal of Agricultural, Biological and Environmental
  Statistics}, 23(2):297--316.

\bibitem[Campbell et~al., 2006]{campbell2006}
Campbell, K., McKay, M.~D., and Williams, B.~J. (2006).
\newblock Sensitivity analysis when model outputs are functions.
\newblock {\em Reliability Engineering \& System Safety}, 91(10):1468--1472.

\bibitem[Chen et~al., 2021]{chen2021}
Chen, J., Mak, S., Joseph, V.~R., and Zhang, C. (2021).
\newblock Function-on-function kriging, with applications to three-dimensional
  printing of aortic tissues.
\newblock {\em Technometrics}, 63(3):384--395.

\bibitem[Cressie, 1993]{cressie1993}
Cressie, N. A.~C. (1993).
\newblock {\em Statistics for Spatial Data}.
\newblock {John Wiley \& Sons, Inc.}

\bibitem[Currin et~al., 1991]{currin1991}
Currin, C., Mitchell, T., Morris, M., and Ylvisaker, D. (1991).
\newblock Bayesian prediction of deterministic functions, with applications to
  the design and analysis of computer experiments.
\newblock {\em Journal of the American Statistical Association},
  86(416):953--963.

\bibitem[Dahl et~al., 2019]{Rxtable2019}
Dahl, D.~B., Scott, D., Roosen, C., Magnusson, A., and Swinton, J. (2019).
\newblock {\em xtable: Export Tables to LaTeX or HTML}.
\newblock R package version 1.8-4.

\bibitem[Damianou and Lawrence, 2013]{damianou2013}
Damianou, A. and Lawrence, N. (2013).
\newblock Deep {{Gaussian}} processes.
\newblock In {\em Artificial {{Intelligence}} and {{Statistics}}}, pages
  207--215. {PMLR}.

\bibitem[{de Jonge}, 2020]{Rdocopt2020}
{de Jonge}, E. (2020).
\newblock {\em docopt: Command-Line Interface Specification Language}.
\newblock R package version 0.7.1.

\bibitem[Dowle and Srinivasan, 2021]{Rdatatable2021}
Dowle, M. and Srinivasan, A. (2021).
\newblock {\em data.table: Extension of `data.frame`}.
\newblock R package version 1.14.2.

\bibitem[Fisher et~al., 2019]{fisher2019}
Fisher, A., Rudin, C., and Dominici, F. (2019).
\newblock All models are wrong, but many are useful: Learning a variable's
  importance by studying an entire class of prediction models simultaneously.
\newblock {\em Journal of machine learning research: JMLR}, 20:177.

\bibitem[Flegal et~al., 2021]{Rmcmcse2021}
Flegal, J.~M., Hughes, J., Vats, D., Dai, N., Gupta, K., and Maji, U. (2021).
\newblock {\em mcmcse: Monte Carlo Standard Errors for MCMC}.
\newblock Riverside, CA, and Kanpur, India.
\newblock R package version 1.5-0.

\bibitem[Frasca, 2019]{Rlogging2019}
Frasca, M. (2019).
\newblock {\em logging: {R} Logging Package}.
\newblock R package version 0.10-108.

\bibitem[Genz et~al., 2021]{Rmvtnorm2021}
Genz, A., Bretz, F., Miwa, T., Mi, X., Leisch, F., Scheipl, F., and Hothorn, T.
  (2021).
\newblock {\em {mvtnorm}: Multivariate Normal and t Distributions}.
\newblock R package version 1.1-3.

\bibitem[Geweke, 1991]{geweke1991}
Geweke, J.~F. (1991).
\newblock Evaluating the accuracy of sampling-based approaches to the
  calculation of posterior moments.
\newblock Technical Report 148, {Federal Reserve Bank of Minneapolis}.

\bibitem[Gneiting and Raftery, 2007]{gneiting2007}
Gneiting, T. and Raftery, A.~E. (2007).
\newblock Strictly proper scoring rules, prediction, and estimation.
\newblock {\em Journal of the American Statistical Association},
  102(477):359--378.

\bibitem[Gramacy, 2020]{gramacy2020}
Gramacy, R.~B. (2020).
\newblock {\em Surrogates: {{Gaussian}} Process Modeling, Design and
  Optimization for the Applied Sciences}.
\newblock {Chapman Hall/CRC}, {Boca Raton, Florida}.

\bibitem[Gramacy and Apley, 2015]{gramacy2015}
Gramacy, R.~B. and Apley, D.~W. (2015).
\newblock Local {{Gaussian}} process approximation for large computer
  experiments.
\newblock {\em Journal of Computational and Graphical Statistics},
  24(2):561--578.

\bibitem[Happ and Greven, 2018]{happ2018}
Happ, C. and Greven, S. (2018).
\newblock Multivariate {{Functional Principal Component Analysis}} for {{Data
  Observed}} on {{Different}} ({{Dimensional}}) {{Domains}}.
\newblock {\em Journal of the American Statistical Association},
  113(522):649--659.

\bibitem[Higdon et~al., 2008]{higdon2008}
Higdon, D., Gattiker, J., Williams, B., and Rightley, M. (2008).
\newblock Computer model calibration using high-dimensional output.
\newblock {\em Journal of the American Statistical Association},
  103(482):570--583.

\bibitem[Hobbs et~al., 2017]{hobbs2017}
Hobbs, J., Braverman, A., Cressie, N., Granat, R., and Gunson, M. (2017).
\newblock Simulation-based uncertainty quantification for estimating
  atmospheric {{CO2}} from satellite data.
\newblock {\em SIAM/ASA Journal on Uncertainty Quantification}, 5(1):956--985.

\bibitem[Hoffman and Gelman, 2014]{hoffman2014}
Hoffman, M.~D. and Gelman, A. (2014).
\newblock The {{No-U-Turn Sampler}}: Adaptively setting path lengths in
  {{Hamiltonian Monte Carlo}}.

\bibitem[Hooker et~al., 2021]{hooker2021}
Hooker, G., Mentch, L., and Zhou, S. (2021).
\newblock Unrestricted permutation forces extrapolation: Variable importance
  requires at least one more model, or there is no free variable importance.
\newblock {\em arXiv:1905.03151 [cs, stat]}.

\bibitem[Iooss and Ribatet, 2009]{iooss2009}
Iooss, B. and Ribatet, M. (2009).
\newblock Global sensitivity analysis of computer models with functional
  inputs.
\newblock {\em Reliability Engineering \& System Safety}, 94(7):1194--1204.

\bibitem[Johnson et~al., 2020]{johnson2020}
Johnson, M., Teixeira, J., Livesey, N., and Braverman, A. (2020).
\newblock Forward model emulation for {{NASA}}'s {{Microwave Limb Sounder}}.
\newblock ASA 2020 Joint Statistical Meetings. Virtual.

\bibitem[Jones et~al., 1998]{jones1998}
Jones, D.~R., Schonlau, M., and Welch, W.~J. (1998).
\newblock Efficient global optimization of expensive black-box functions.
\newblock {\em Journal of Global Optimization}, 13(4):455--492.

\bibitem[Katzfuss and Guinness, 2021]{katzfuss2021}
Katzfuss, M. and Guinness, J. (2021).
\newblock A general framework for vecchia approximations of gaussian processes.
\newblock {\em Statistical Science}, 36(1).

\bibitem[Kennedy and O'Hagan, 2001]{kennedy2001}
Kennedy, M.~C. and O'Hagan, A. (2001).
\newblock Bayesian calibration of computer models.
\newblock {\em Journal of the Royal Statistical Society: Series B (Statistical
  Methodology)}, 63(3):425--464.

\bibitem[Koehler and Owen, 1996]{koehler1996}
Koehler, J. and Owen, A. (1996).
\newblock 9 {{Computer}} experiments.
\newblock In {\em Handbook of {{Statistics}}}, volume~13, pages 261--308.
  {Elsevier}.

\bibitem[Kuttubekova, 2019]{kuttubekova2019}
Kuttubekova, G. (2019).
\newblock Emulator for water erosion prediction project computer model using
  gaussian processes with functional inputs.
\newblock {\em Creative Components}.

\bibitem[Lamminp{\"a}{\"a} et~al., 2019]{lamminpaa2019}
Lamminp{\"a}{\"a}, O., Hobbs, J., Brynjarsd{\'o}ttir, J., Laine, M., Braverman,
  A., Lindqvist, H., and Tamminen, J. (2019).
\newblock Accelerated {{MCMC}} for {{Satellite-Based Measurements}} of
  {{Atmospheric CO2}}.
\newblock {\em Remote Sensing}, 11(17):2061.

\bibitem[Li and Tan, 2021]{li2021}
Li, Z. and Tan, M. H.~Y. (2021).
\newblock A {{Gaussian}} process emulator based approach for {{Bayesian}}
  calibration of a functional input.
\newblock {\em Technometrics}, pages 1--13.

\bibitem[Livesey et~al., 2006]{livesey2006}
Livesey, N., Van~Snyder, W., Read, W., and Wagner, P. (2006).
\newblock Retrieval algorithms for the {{EOS Microwave Limb Sounder}}
  ({{MLS}}).
\newblock {\em IEEE Transactions on Geoscience and Remote Sensing},
  44(5):1144--1155.

\bibitem[Livesey et~al., 2020]{liversey2020}
Livesey, N.~J., Read, W.~G., Wagner, P.~A., Froidevaux, L., Santee, M.~L.,
  Schwartz, M.~J., Lambert, A., Valle, L. F.~M., Pumphrey, H.~C., Manney,
  G.~L., Fuller, R.~A., Jarnot, R.~F., Knosp, B.~W., and Lay, R.~R. (2020).
\newblock Earth {{Observing System}} ({{EOS}}) {{Aura Microwave Limb Sounder}}
  ({{MLS}}) version 5.0x level 2 and 3 data quality and description document.

\bibitem[Ma et~al., 2021]{ma2021}
Ma, P., Mondal, A., Konomi, B.~A., Hobbs, J., Song, J.~J., and Kang, E.~L.
  (2021).
\newblock Computer model emulation with high-dimensional functional output in
  large-scale observing system uncertainty experiments.
\newblock {\em Technometrics}, 0(0):1--15.

\bibitem[MacKay, 1996]{mackay1996}
MacKay, D. J.~C. (1996).
\newblock Bayesian non-linear modeling for the prediction competition.
\newblock In Heidbreder, G.~R., editor, {\em Maximum {{Entropy}} and {{Bayesian
  Methods}}}, pages 221--234. {Springer Netherlands}, {Dordrecht}.

\bibitem[Morris, 2012]{morris2012}
Morris, M.~D. (2012).
\newblock Gaussian surrogates for computer models with time-varying inputs and
  outputs.
\newblock {\em Technometrics}, 54(1):42--50.

\bibitem[Muehlenstaedt et~al., 2017]{muehlenstaedt2017}
Muehlenstaedt, T., Fruth, J., and Roustant, O. (2017).
\newblock Computer experiments with functional inputs and scalar outputs by a
  norm-based approach.
\newblock {\em Statistics and Computing}, 27(4):1083--1097.

\bibitem[Nanty et~al., 2016]{nanty2016}
Nanty, S., Helbert, C., Marrel, A., P{\'e}rot, N., and Prieur, C. (2016).
\newblock Sampling, metamodeling, and sensitivity analysis of numerical
  simulators with functional stochastic inputs.
\newblock {\em SIAM/ASA Journal on Uncertainty Quantification}, 4(1):636--659.

\bibitem[Neal, 1996]{neal1996}
Neal, R.~M. (1996).
\newblock {\em Bayesian Learning for Neural Networks}, volume 118 of {\em
  Lecture {{Notes}} in {{Statistics}}}.
\newblock {Springer New York}, {New York, NY}.

\bibitem[Nicodemus et~al., 2010]{nicodemus2010}
Nicodemus, K.~K., Malley, J.~D., Strobl, C., and Ziegler, A. (2010).
\newblock The behaviour of random forest permutation-based variable importance
  measures under predictor correlation.
\newblock {\em BMC Bioinformatics}, 11(1):110.

\bibitem[O'Hagan, 1992]{ohagan1992}
O'Hagan, A. (1992).
\newblock Some {{Bayesian}} numerical analysis.
\newblock In Bernardo, J.~M., editor, {\em Bayesian Statistics. 4}, pages
  345--363. {Oxford University Press}.

\bibitem[Piironen and Vehtari, 2016]{piironen2016}
Piironen, J. and Vehtari, A. (2016).
\newblock Projection predictive model selection for {{Gaussian}} processes.
\newblock In {\em 2016 {{IEEE}} 26th {{International Workshop}} on {{Machine
  Learning}} for {{Signal Processing}} ({{MLSP}})}, pages 1--6.

\bibitem[{R Core Team}, 2021]{Rcore2021}
{R Core Team} (2021).
\newblock {\em R: A Language and Environment for Statistical Computing}.
\newblock R Foundation for Statistical Computing, Vienna, Austria.

\bibitem[Raftery and Lewis, 1992]{raftery1992}
Raftery, A.~E. and Lewis, S.~M. (1992).
\newblock [{{Practical Markov}} chain {{Monte Carlo}}]: {{Comment}}: One long
  run with diagnostics: Implementation strategies for {{Markov}} chain {{Monte
  Carlo}}.
\newblock {\em Statistical Science}, 7(4).

\bibitem[Ramsay et~al., 2021]{Rfda2021}
Ramsay, J.~O., Graves, S., and Hooker, G. (2021).
\newblock {\em fda: Functional Data Analysis}.
\newblock R package version 5.5.1.

\bibitem[Ramsay and Silverman, 2005]{ramsay2005}
Ramsay, J.~O. and Silverman, B.~W. (2005).
\newblock {\em Functional Data Analysis}.
\newblock Springer {{Series}} in {{Statistics}}. {Springer New York}, {New
  York, NY}.

\bibitem[Read et~al., 2006]{read2006}
Read, W., Shippony, Z., Schwartz, M., Livesey, N., and Van~Snyder, W. (2006).
\newblock The clear-sky unpolarized forward model for the {{EOS}} aura
  {{Microwave Limb Sounder}} ({{MLS}}).
\newblock {\em IEEE Transactions on Geoscience and Remote Sensing},
  44(5):1367--1379.

\bibitem[Sacks et~al., 1989a]{sacks1989a}
Sacks, J., Schiller, S.~B., and Welch, W.~J. (1989a).
\newblock Designs for computer experiments.
\newblock {\em Technometrics}, 31(1):41--47.

\bibitem[Sacks et~al., 1989b]{sacks1989}
Sacks, J., Welch, W.~J., Mitchell, T.~J., and Wynn, H.~P. (1989b).
\newblock Design and analysis of computer experiments.
\newblock {\em Statistical Science}, 4(4):409--423.

\bibitem[Santner et~al., 2018]{santner2018}
Santner, T.~J., Williams, B.~J., and Notz, W.~I. (2018).
\newblock {\em The Design and Analysis of Computer Experiments}.
\newblock Springer {{Series}} in {{Statistics}}. {Springer New York}, {New
  York, NY}.

\bibitem[Sauer et~al., 2022a]{sauer2022}
Sauer, A., Cooper, A., and Gramacy, R.~B. (2022a).
\newblock Vecchia-approximated {{Deep Gaussian}} processes for computer
  experiments.

\bibitem[Sauer et~al., 2022b]{sauer2022a}
Sauer, A., Gramacy, R.~B., and Higdon, D. (2022b).
\newblock Active learning for {{Deep Gaussian}} process surrogates.
\newblock {\em Technometrics}, pages 1--15.

\bibitem[Schloerke et~al., 2021]{RGGally2021}
Schloerke, B., Cook, D., Larmarange, J., Briatte, F., Marbach, M., Thoen, E.,
  Elberg, A., and Crowley, J. (2021).
\newblock {\em GGally: Extension to 'ggplot2'}.
\newblock R package version 2.1.2.

\bibitem[Schwartz et~al., 2006]{schwartz2006}
Schwartz, M., Read, W., and Van~Snyder, W. (2006).
\newblock {{EOS MLS}} forward model polarized radiative transfer for
  zeeman-split oxygen lines.
\newblock {\em IEEE Transactions on Geoscience and Remote Sensing},
  44(5):1182--1191.

\bibitem[Solymos and Zawadzki, 2021]{Rpbapply2021}
Solymos, P. and Zawadzki, Z. (2021).
\newblock {\em pbapply: Adding Progress Bar to '*apply' Functions}.
\newblock R package version 1.5-0.

\bibitem[{Stan Development Team}, 2020]{Rrstan2212}
{Stan Development Team} (2020).
\newblock {RStan}: the {R} interface to {Stan}.
\newblock R package version 2.21.2.

\bibitem[{Stan Development Team}, 2021]{Stan221}
{Stan Development Team} (2021).
\newblock Stan modeling language users guide and reference manual.
\newblock 2.21.

\bibitem[Strobl et~al., 2008]{strobl2008}
Strobl, C., Boulesteix, A.-L., Kneib, T., Augustin, T., and Zeileis, A. (2008).
\newblock Conditional variable importance for random forests.
\newblock {\em BMC Bioinformatics}, 9(1):307.

\bibitem[Strobl et~al., 2007]{strobl2007}
Strobl, C., Boulesteix, A.-L., Zeileis, A., and Hothorn, T. (2007).
\newblock Bias in random forest variable importance measures: Illustrations,
  sources and a solution.
\newblock {\em BMC Bioinformatics}, 8(1):25.

\bibitem[Tan and Li, 2019]{tan2019}
Tan, M. H.~Y. and Li, G. (2019).
\newblock Gaussian process modeling using the principle of superposition.
\newblock {\em Technometrics}, 61(2):202--218.

\bibitem[Turmon and Braverman, 2019]{turmon2019}
Turmon, M. and Braverman, A. (2019).
\newblock Uncertainty quantification for {{JPL}} retrievals.
\newblock Technical {{Report}}, {Pasadena, CA: Jet Propulsion Laboratory,
  National Aeronautics and Space Administration, 2019}.

\bibitem[{van den Brand}, 2021]{Rggh4x2021}
{van den Brand}, T. (2021).
\newblock {\em ggh4x: Hacks for 'ggplot2'}.
\newblock R package version 0.2.1.

\bibitem[{van der Loo}, 2020]{Rtinytest2020}
{van der Loo}, M. (2020).
\newblock A method for deriving information from running {R} code.
\newblock {\em The {R} Journal}, page Accepted for publication.

\bibitem[Vecchia, 1988]{vecchia1988}
Vecchia, A.~V. (1988).
\newblock Estimation and model identification for continuous spatial processes.
\newblock {\em Journal of the Royal Statistical Society. Series B
  (Methodological)}, 50(2):297--312.

\bibitem[Wang et~al., 2017]{wang2017}
Wang, B., Chen, T., and Xu, A. (2017).
\newblock Gaussian process regression with functional covariates and
  multivariate response.
\newblock {\em Chemometrics and Intelligent Laboratory Systems}, 163:1--6.

\bibitem[Wang and Xu, 2019]{wang2019}
Wang, B. and Xu, A. (2019).
\newblock Gaussian process methods for nonparametric functional regression with
  mixed predictors.
\newblock {\em Computational Statistics \& Data Analysis}, 131:80--90.

\bibitem[Waters et~al., 2006]{waters2006}
Waters, J., Froidevaux, L., Harwood, R., Jarnot, R., Pickett, H., Read, W.,
  Siegel, P., Cofield, R., Filipiak, M., Flower, D., Holden, J., Lau, G.,
  Livesey, N., Manney, G., Pumphrey, H., Santee, M., Wu, D., Cuddy, D., Lay,
  R., Loo, M., Perun, V., Schwartz, M., Stek, P., Thurstans, R., Boyles, M.,
  Chandra, K., Chavez, M., {Gun-Shing Chen}, Chudasama, B., Dodge, R., Fuller,
  R., Girard, M., Jiang, J., {Yibo Jiang}, Knosp, B., LaBelle, R., Lam, J.,
  Lee, K., Miller, D., Oswald, J., Patel, N., Pukala, D., Quintero, O., Scaff,
  D., Van~Snyder, W., Tope, M., Wagner, P., and Walch, M. (2006).
\newblock The earth observing system {{Microwave Limb Sounder}} ({{EOS MLS}})
  on the {{Aura}} satellite.
\newblock {\em IEEE Transactions on Geoscience and Remote Sensing},
  44(5):1075--1092.

\bibitem[Waters et~al., 1999]{waters1999}
Waters, J.~W., Read, W.~G., Froidevaux, L., Jarnot, R.~F., Cofield, R.~E.,
  Flower, D.~A., Lau, G.~K., Pickett, H.~M., Santee, M.~L., Wu, D.~L., Boyles,
  M.~A., Burke, J.~R., Lay, R.~R., Loo, M.~S., Livesey, N.~J., Lungu, T.~A.,
  Manney, G.~L., Nakamura, L.~L., Perun, V.~S., Ridenoure, B.~P., Shippony, Z.,
  Siegel, P.~H., Thurstans, R.~P., Harwood, R.~S., Pumphrey, H.~C., and
  Filipiak, M.~J. (1999).
\newblock The {{UARS}} and {{EOS Microwave Limb Sounder}} ({{MLS}})
  experiments.
\newblock {\em Journal of the Atmospheric Sciences}, 56(2):194--218.

\bibitem[Wickham, 2016]{Rggplot2}
Wickham, H. (2016).
\newblock {\em ggplot2: Elegant Graphics for Data Analysis}.
\newblock Springer-Verlag New York.

\bibitem[Wipf and Nagarajan, 2007]{wipf2007}
Wipf, D. and Nagarajan, S. (2007).
\newblock A new view of automatic relevance determination.
\newblock In Platt, J., Koller, D., Singer, Y., and Roweis, S., editors, {\em
  Advances in Neural Information Processing Systems}, volume~20. {Curran
  Associates, Inc.}

\end{thebibliography}


\begin{thebibliography}{10}

\bibitem{RgridExtra2017}
Baptiste Auguie.
\newblock {\em gridExtra: Miscellaneous Functions for "Grid" Graphics}, 2017.
\newblock R package version 2.3.

\bibitem{barath1993}
F.~T. Barath, M.~C. Chavez, R.~E. Cofield, D.~A. Flower, M.~A. Frerking, M.~B.
  Gram, W.~M. Harris, J.~R. Holden, R.~F. Jarnot, W.~G. Kloezeman, G.~J. Klose,
  G.~K. Lau, M.~S. Loo, B.~J. Maddison, R.~J. Mattauch, R.~P. McKinney, G.~E.
  Peckham, H.~M. Pickett, G.~Siebes, F.~S. Soltis, R.~A. Suttie, J.~A. Tarsala,
  J.~W. Waters, and W.~J. Wilson.
\newblock The {{Upper Atmosphere Research Satellite}} microwave limb sounder
  instrument.
\newblock {\em Journal of Geophysical Research}, 98(D6):10751, 1993.

\bibitem{bastos2009}
Leonardo~S. Bastos and Anthony O'Hagan.
\newblock Diagnostics for {{Gaussian Process Emulators}}.
\newblock {\em Technometrics}, 51(4):425--438, November 2009.

\bibitem{bayarri2007a}
M.~J. Bayarri, D.~Walsh, J.~O. Berger, J.~Cafeo, G.~{Garcia-Donato}, F.~Liu,
  J.~Palomo, R.~J. Parthasarathy, R.~Paulo, and J.~Sacks.
\newblock Computer model validation with functional output.
\newblock {\em The Annals of Statistics}, 35(5):1874--1906, October 2007.

\bibitem{betancourt2016}
Michael Betancourt.
\newblock Diagnosing {{Suboptimal Cotangent Disintegrations}} in {{Hamiltonian
  Monte Carlo}}.
\newblock {\em arXiv:1604.00695 [stat]}, April 2016.

\bibitem{campbell2006}
Katherine Campbell, Michael~D. McKay, and Brian~J. Williams.
\newblock Sensitivity analysis when model outputs are functions.
\newblock {\em Reliability Engineering \& System Safety}, 91(10):1468--1472,
  October 2006.

\bibitem{currin1991}
Carla Currin, Toby Mitchell, Max Morris, and Don Ylvisaker.
\newblock Bayesian {{Prediction}} of {{Deterministic Functions}}, with
  {{Applications}} to the {{Design}} and {{Analysis}} of {{Computer
  Experiments}}.
\newblock {\em Journal of the American Statistical Association},
  86(416):953--963, December 1991.

\bibitem{Rxtable2019}
David~B. Dahl, David Scott, Charles Roosen, Arni Magnusson, and Jonathan
  Swinton.
\newblock {\em xtable: Export Tables to LaTeX or HTML}, 2019.
\newblock R package version 1.8-4.

\bibitem{damianou2013}
Andreas Damianou and Neil Lawrence.
\newblock Deep {{Gaussian Processes}}.
\newblock In {\em Artificial {{Intelligence}} and {{Statistics}}}, pages
  207--215. {PMLR}, April 2013.

\bibitem{datta2016}
Abhirup Datta, Sudipto Banerjee, Andrew~O. Finley, and Alan~E. Gelfand.
\newblock Hierarchical {{Nearest-Neighbor Gaussian Process Models}} for {{Large
  Geostatistical Datasets}}.
\newblock {\em Journal of the American Statistical Association},
  111(514):800--812, April 2016.

\bibitem{Rdocopt2020}
Edwin {de Jonge}.
\newblock {\em docopt: Command-Line Interface Specification Language}, 2020.
\newblock R package version 0.7.1.

\bibitem{Rdatatable2021}
Matt Dowle and Arun Srinivasan.
\newblock {\em data.table: Extension of `data.frame`}, 2021.
\newblock R package version 1.14.2.

\bibitem{Rmcmcse2021}
James~M. Flegal, John Hughes, Dootika Vats, Ning Dai, Kushagra Gupta, and
  Uttiya Maji.
\newblock {\em mcmcse: Monte Carlo Standard Errors for MCMC}.
\newblock Riverside, CA, and Kanpur, India, 2021.
\newblock R package version 1.5-0.

\bibitem{Rlogging2019}
Mario Frasca.
\newblock {\em logging: R Logging Package}, 2019.
\newblock R package version 0.10-108.

\bibitem{Rmvtnorm2021}
Alan Genz, Frank Bretz, Tetsuhisa Miwa, Xuefei Mi, Friedrich Leisch, Fabian
  Scheipl, and Torsten Hothorn.
\newblock {\em {mvtnorm}: Multivariate Normal and t Distributions}, 2021.
\newblock R package version 1.1-3.

\bibitem{gneiting2007}
Tilmann Gneiting and Adrian~E Raftery.
\newblock Strictly {{Proper Scoring Rules}}, {{Prediction}}, and
  {{Estimation}}.
\newblock {\em Journal of the American Statistical Association},
  102(477):359--378, March 2007.

\bibitem{gramacy2020}
Robert~B. Gramacy.
\newblock {\em Surrogates: {{Gaussian}} Process Modeling, Design and
  Optimization for the Applied Sciences}.
\newblock {Chapman Hall/CRC}, {Boca Raton, Florida}, 2020.

\bibitem{gramacy2015}
Robert~B. Gramacy and Daniel~W. Apley.
\newblock Local {{Gaussian Process Approximation}} for {{Large Computer
  Experiments}}.
\newblock {\em Journal of Computational and Graphical Statistics},
  24(2):561--578, April 2015.

\bibitem{higdon2008}
Dave Higdon, James Gattiker, Brian Williams, and Maria Rightley.
\newblock Computer {{Model Calibration Using High-Dimensional Output}}.
\newblock {\em Journal of the American Statistical Association},
  103(482):570--583, June 2008.

\bibitem{hobbs2017}
Jonathan Hobbs, Amy Braverman, Noel Cressie, Robert Granat, and Michael Gunson.
\newblock Simulation-{{Based Uncertainty Quantification}} for {{Estimating
  Atmospheric CO}}\$\_2\$ from {{Satellite Data}}.
\newblock {\em SIAM/ASA Journal on Uncertainty Quantification}, 5(1):956--985,
  January 2017.

\bibitem{hoffman2014}
Matthew~D. Hoffman and Andrew Gelman.
\newblock The {{No-U-Turn Sampler}}: {{Adaptively Setting Path Lengths}} in
  {{Hamiltonian Monte Carlo}}, 2014.

\bibitem{johnson2020}
Margaret Johnson.
\newblock Forward {{Model Emulation}} for {{NASA}}'s {{Microwave Limb
  Sounder}}, August 2020.

\bibitem{jones1998}
Donald~R. Jones, Matthias Schonlau, and William~J. Welch.
\newblock Efficient {{Global Optimization}} of {{Expensive Black-Box
  Functions}}.
\newblock {\em Journal of Global Optimization}, 13(4):455--492, December 1998.

\bibitem{kennedy2001}
Marc~C. Kennedy and Anthony O'Hagan.
\newblock Bayesian calibration of computer models.
\newblock {\em Journal of the Royal Statistical Society: Series B (Statistical
  Methodology)}, 63(3):425--464, 2001.

\bibitem{koehler1996}
J.R. Koehler and A.B. Owen.
\newblock 9 {{Computer}} experiments.
\newblock In {\em Handbook of {{Statistics}}}, volume~13, pages 261--308.
  {Elsevier}, 1996.

\bibitem{ma2021}
Pulong Ma, Anirban Mondal, Bledar~A. Konomi, Jonathan Hobbs, Joon~Jin Song, and
  Emily~L. Kang.
\newblock Computer {{Model Emulation}} with {{High-Dimensional Functional
  Output}} in {{Large-Scale Observing System Uncertainty Experiments}}.
\newblock {\em Technometrics}, 0(0):1--15, March 2021.

\bibitem{morris2012}
Max~D. Morris.
\newblock Gaussian {{Surrogates}} for {{Computer Models With Time-Varying
  Inputs}} and {{Outputs}}.
\newblock {\em Technometrics}, 54(1):42--50, February 2012.

\bibitem{muehlenstaedt2017}
Thomas Muehlenstaedt, Jana Fruth, and Olivier Roustant.
\newblock Computer experiments with functional inputs and scalar outputs by a
  norm-based approach.
\newblock {\em Statistics and Computing}, 27(4):1083--1097, July 2017.

\bibitem{muehlenstaedt2017a}
Thomas Muehlenstaedt, Jana Fruth, and Olivier Roustant.
\newblock Computer experiments with functional inputs and scalar outputs by a
  norm-based approach.
\newblock {\em Statistics and Computing}, 27(4):1083--1097, July 2017.

\bibitem{nanty2016}
Simon Nanty, C{\'e}line Helbert, Amandine Marrel, Nadia P{\'e}rot, and
  Cl{\'e}mentine Prieur.
\newblock Sampling, {{Metamodeling}}, and {{Sensitivity Analysis}} of
  {{Numerical Simulators}} with {{Functional Stochastic Inputs}}.
\newblock {\em SIAM/ASA Journal on Uncertainty Quantification}, 4(1):636--659,
  January 2016.

\bibitem{liversey2020}
{Nathaniel J. Livesey}, {William G. Read}, {Paul A. Wagner}, {Lucien
  Froidevaux}, {Michelle L. Santee}, {Michael J. Schwartz}, {Alyn Lambert},
  {Luis F. Mill\'an Valle}, {Hugh C. Pumphrey}, {Gloria L. Manney}, {Ryan A.
  Fuller}, {Robert F. Jarnot}, {Brian W. Knosp}, and {Richard R. Lay}.
\newblock Earth {{Observing System}} ({{EOS}}) {{Aura Microwave Limb Sounder}}
  ({{MLS}}) {{Version}} 5.0x {{Level}} 2 and 3 data quality and description
  document, June 2020.

\bibitem{neal1996}
Radford~M. Neal.
\newblock {\em Bayesian {{Learning}} for {{Neural Networks}}}, volume 118 of
  {\em Lecture {{Notes}} in {{Statistics}}}.
\newblock {Springer New York}, {New York, NY}, 1996.

\bibitem{ohagan1992}
A.~O'Hagan.
\newblock Some {{Bayesian}} numerical analysis.
\newblock In Jos{\'e}~M. Bernardo, editor, {\em Bayesian Statistics. 4}, pages
  345--363. {Oxford University Press}, 1992.

\bibitem{piironen2016}
Juho Piironen and Aki Vehtari.
\newblock Projection predictive model selection for {{Gaussian}} processes.
\newblock In {\em 2016 {{IEEE}} 26th {{International Workshop}} on {{Machine
  Learning}} for {{Signal Processing}} ({{MLSP}})}, pages 1--6, September 2016.

\bibitem{Rcore2021}
{R Core Team}.
\newblock {\em R: A Language and Environment for Statistical Computing}.
\newblock R Foundation for Statistical Computing, Vienna, Austria, 2021.

\bibitem{Rparallel}
{R Core Team}.
\newblock {\em R: A Language and Environment for Statistical Computing}.
\newblock R Foundation for Statistical Computing, Vienna, Austria, 2021.

\bibitem{Rfda2021}
J.~O. Ramsay, Spencer Graves, and Giles Hooker.
\newblock {\em fda: Functional Data Analysis}, 2021.
\newblock R package version 5.5.1.

\bibitem{ramsay2005}
J.~O. Ramsay and B.~W. Silverman.
\newblock {\em Functional {{Data Analysis}}}.
\newblock Springer {{Series}} in {{Statistics}}. {Springer New York}, {New
  York, NY}, 2005.

\bibitem{sacks1989a}
Jerome Sacks, Susannah~B. Schiller, and William~J. Welch.
\newblock Designs for {{Computer Experiments}}.
\newblock {\em Technometrics}, 31(1):41--47, 1989.

\bibitem{sacks1989}
Jerome Sacks, William~J. Welch, Toby~J. Mitchell, and Henry~P. Wynn.
\newblock Design and {{Analysis}} of {{Computer Experiments}}.
\newblock {\em Statistical Science}, 4(4):409--423, November 1989.

\bibitem{santner2018h}
Thomas~J. Santner, Brian~J. Williams, and William~I. Notz.
\newblock {\em The {{Design}} and {{Analysis}} of {{Computer Experiments}}}.
\newblock Springer {{Series}} in {{Statistics}}. {Springer New York : Imprint:
  Springer}, {New York, NY}, 2nd ed. 2018 edition, 2018.

\bibitem{RGGally2021}
Barret Schloerke, Di~Cook, Joseph Larmarange, Francois Briatte, Moritz Marbach,
  Edwin Thoen, Amos Elberg, and Jason Crowley.
\newblock {\em GGally: Extension to 'ggplot2'}, 2021.
\newblock R package version 2.1.2.

\bibitem{Rpbapply2021}
Peter Solymos and Zygmunt Zawadzki.
\newblock {\em pbapply: Adding Progress Bar to '*apply' Functions}, 2021.
\newblock R package version 1.5-0.

\bibitem{Rrstan2212}
{Stan Development Team}.
\newblock {RStan}: the {R} interface to {Stan}, 2020.
\newblock R package version 2.21.2.

\bibitem{Stan221}
{Stan Development Team}.
\newblock Stan modeling language users guide and reference manual, 2021.
\newblock 2.21.

\bibitem{turmon2019a}
Michael Turmon and Amy Braverman.
\newblock Uncertainty quantification for {{JPL}} retrievals.
\newblock Technical {{Report}}, {Pasadena, CA: Jet Propulsion Laboratory,
  National Aeronautics and Space Administration, 2019}, February 2019.

\bibitem{Rggh4x2021}
Teun {van den Brand}.
\newblock {\em ggh4x: Hacks for 'ggplot2'}, 2021.
\newblock R package version 0.2.1.

\bibitem{Rtinytest2020}
MPJ {van der Loo}.
\newblock A method for deriving information from running r code.
\newblock {\em The R Journal}, page Accepted for publication, 2020.

\bibitem{wang2019}
Bo~Wang and Aiping Xu.
\newblock Gaussian process methods for nonparametric functional regression with
  mixed predictors.
\newblock {\em Computational Statistics \& Data Analysis}, 131:80--90, March
  2019.

\bibitem{Rggplot2}
Hadley Wickham.
\newblock {\em ggplot2: Elegant Graphics for Data Analysis}.
\newblock Springer-Verlag New York, 2016.

\end{thebibliography}


\begin{thebibliography}{}

\end{thebibliography}

\end{document}


\def\spacingset#1{\renewcommand{\baselinestretch}%
{#1}\small\normalsize} \spacingset{1}


\author[1]{Luis Damiano}
\author[2]{Margaret Johnson}
\author[2]{Joaquim Teixeira}
\author[3]{Max D. Morris}
\author[1]{Jarad Niemi}
\affil[1]{\small Department of Statistics, Iowa State University, Ames, IA, US}
\affil[2]{\small Jet Propulsion Laboratory, California Institute of Technology,
  Pasadena, CA, US}
\affil[3]{\small Departments of Statistics, and Industrial and Manufacturing
Systems Engineering, Iowa State University, Ames, IA, US}

\if0\blind{}
{
  \title{\bf
    Automatic Dynamic Relevance Determination \\
    for Gaussian process regression \\
    with high-dimensional functional inputs
  }
  \maketitle
} \fi

\if1\blind{}
{
  \bigskip
  \bigskip
  \bigskip
  \begin{center}
    { \LARGE\bf
      Automatic Dynamic Relevance Determination \\
      for Gaussian process regression \\
      with high-dimensional functional inputs }
  \end{center}
  \medskip
} \fi
\spacingset{2}

\begin{center}
{\large\bf SUPPLEMENTARY MATERIAL}
\end{center}

{
  \renewcommand{\thesection}{S\arabic{section}}
  \renewcommand{\thetable}{S\arabic{table}}
  \renewcommand{\thefigure}{S\arabic{figure}}
  \section{Nomenclature}%
\label{app:nomenclature}

\begin{center}
  \begin{tabular}{rll}
    \toprule
    \multicolumn{3}{l}{\textsc{Indices for finite sequences}}
    \\
    $h$ &= $1, \dots, H \in \mathbb{N}$
    & Training/test subset \\
    $i, j$ &= $1, \dots, N \in \mathbb{N}$
    & Profiles \\
    $k_q $ &= $ 1, \dots, K_q = \mathbb{N}$
    & Measurement per profile (functional evaluation) \\
    ${\tilde{k}}_q $ &= $ 1, \dots, \tilde{K}_q \le K_q \in \mathbb{N}$
    & Principal component \\
    $m $ &= $ 1, \dots, M \in \mathbb{N}$
    & Posterior sample \\
    ${\tilde{m}} $ &= $ 1, \dots, \tilde{M} \le M \in \mathbb{N}$
    & Posterior sample after thinning \\
    $p $ &= $ 1, \dots, P = 7$
    & Plausible models \\
    $q $ &= $ 1, \dots, Q = 5$
    & Input variable \\
    $n $ &= $ 1, \dots, N \in \mathbb{N}$
    & Sounding (input-output pair) \\
    \midrule
    \multicolumn{3}{l}{\textsc{Vectors and matrices}}
    \\
    $\bm{y} $ &= $ \left\{y_n\right\}$
    & Output vector \\
    $\bm{t}^{(q)} $ &= $ \left\{t^{(q)}_k\right\}$
    & Index vector \\
    $\bm{X}^{(q)} $ &= $ \left\{x_{n,k}^{(q)}\right\}$
    & Input matrix \\
    $\tilde{\bm{X}}^{(q)} $ &= $ \left\{{\tilde{x}}_{n,{\tilde{k}}}^{(q)}\right\}$
    & Matrix with input principal component scores \\
    \midrule
    \multicolumn{3}{l}{\textsc{Collections}}
    \\
    $\mathcal{D} $ &= $ (\bm{y}, \bm{X})$
    & Training data set \\
    $\mathcal{D}_* $ &= $ (\bm{y}_*, \bm{X}_*)$
    & Test data set \\
   $\mathcal{V} $ &= $\left\{\hat{v}^{(h, \tilde{m}, p, q)}
\right\}_{h, \tilde{m}, p, q}$
    & Validation statistic set \\
    \bottomrule
  \end{tabular}
\end{center}

\section{Data transformation}%
\label{app:data-transformation}

\begin{table}[H]
  \centering
    \begin{tabular}{llrrrrr}
      \toprule
      Input & Unit   & $K$ & $X_l$ & $X_u$ & $t_l$ & $t_u$ \\
      \midrule
      H2O   & log ppm    & 42 & -16.12e-00 & -5.16e-00 & -2.50 & 2.67 \\
      HNO3  & ppm        & 16 &  -7.20e-09 &  1.83e-08 & -2.50 & 0.00 \\
      N2O   & ppm        & 18 &  -4.00e-08 &  6.22e-07 & -2.00 & 0.84 \\
      O3    & ppm        & 39 &  -3.96e-06 &  1.21e-05 & -2.50 & 1.67 \\
      Temp  & log Kelvin & 43 &   4.59e-00 &  5.73e-00 & -2.50 & 3.00 \\
      \bottomrule
    \end{tabular}
  \caption{Measurement units and scaling boundaries.}%
  \label{tab:input-scales}
\end{table}

\section{Validation statistics}%
\label{app:validation}

Let
$\hat{\mathbf{m}} = \predmean = \{\hat{m}_{*n}: n = 1, \dots, N\} =
\EE{\mathbf{y}_* | \mathbf{y}, \mathbf{X}, \mathbf{X}_*}$
and
$\hat{\mathbf{S}} = \predvar = \VV{\mathbf{y}_* |
  \mathbf{y}, \mathbf{X}, \mathbf{X}_*}$
be the predictive mean vector and covariance matrix.
Define the
prediction error vector $\mathbf{e} = \mathbf{e}_{*}^{y} =
\mathbf{y}_{*} - \hat{\mathbf{m}}$, the square Mahalanobis distance $D^2
= \mathbf{e}^{\transp} \hat{\mathbf{S}}^{-1} \mathbf{e}$,
and the point-wise 95\% coverage indicator variable
$I_{n} = 1$ if $y_{*n} \in \hat{m}_{*n} \pm 1.96
{\hat{S}_{nn}}^{-\frac{1}{2}}$ or zero otherwise.
Let $\bar{y}_* = N^{-1} \sum_{n=1}^{N} y_{*n}$ be the test output
mean.
\begin{center}
  \begin{tabular}{lrl}
    RMSE
    & $v_{\textsc{RMSE}}$ =
    & $N^{-\frac{1}{2}} \norm{\mathbf{e}}$ \\
    $R^2$
    & $v_{\textsc{R2}} $ =
    & $1 -%
      \norm{\mathbf{e}}^{2}
      \norm{\mathbf{y}_* - \bar{y}_*}^{-2}$ \\
    PPLD
    & $v_{\textsc{PPLD}}$ =
    & $
      -\frac{1}{2} \log \lvert \hat{\mathbf{S}} \rvert
      -D^2
      -\frac{n}{2} \log 2 \pi
      $
    \\
    CRPS
    & $v_{\textsc{CRPS}}$ =
    & $
      -\log \lvert \hat{\mathbf{S}} \rvert%
      -D^2
      $
    \\
    Nominal coverage
    & $v_{\textsc{COV95}}$ =
    & $N^{-1} \sum_{n = 1}^{N} I_{n}$
  \end{tabular}
\end{center}

\spacingset{1}






\begin{landscape}
  \section{Validation statistic summary for each subset}%

  \begin{figure}[H]
    \centering
    \includegraphics[width=.7\linewidth]{%
      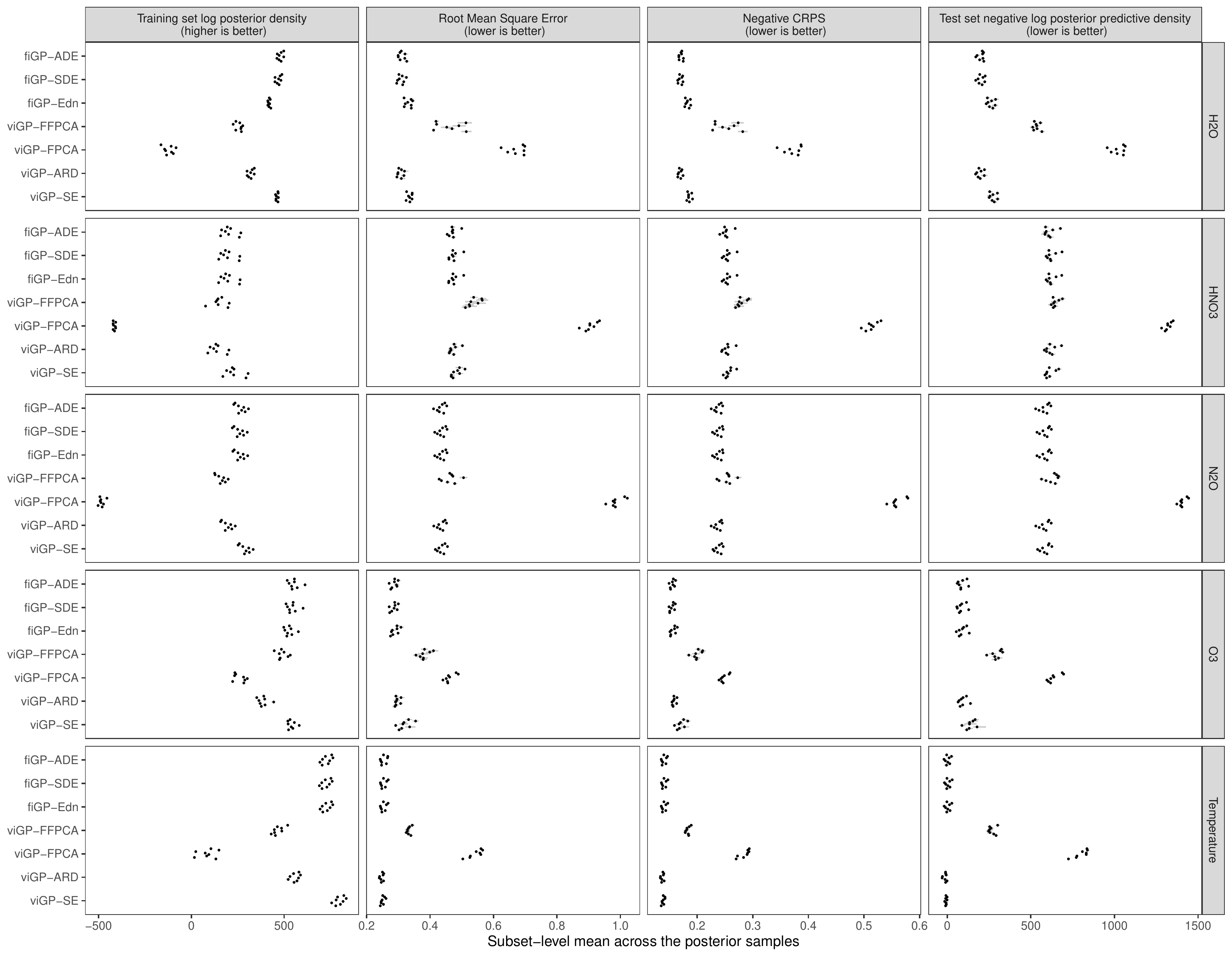}
    \caption{
      Each point corresponds to a subset-level mean statistic across
      the posterior samples $\bar{v}^{(h, p, q)}$ for each plausible model
      $p$ and input variable $q$. Gray lines are 95\% empirical interval of
      ${v}^{(h, \tilde{m}, p, q)}$ across the posterior samples $\tilde{m}$.
    }\label{fig:validation-foldmeans-brief}
  \end{figure}

  \clearpage

  \section{Parameter posterior distributions}%
  \label{app:param-posterior}

  \begin{figure}[H]
    \includegraphics[width=\linewidth]{%
      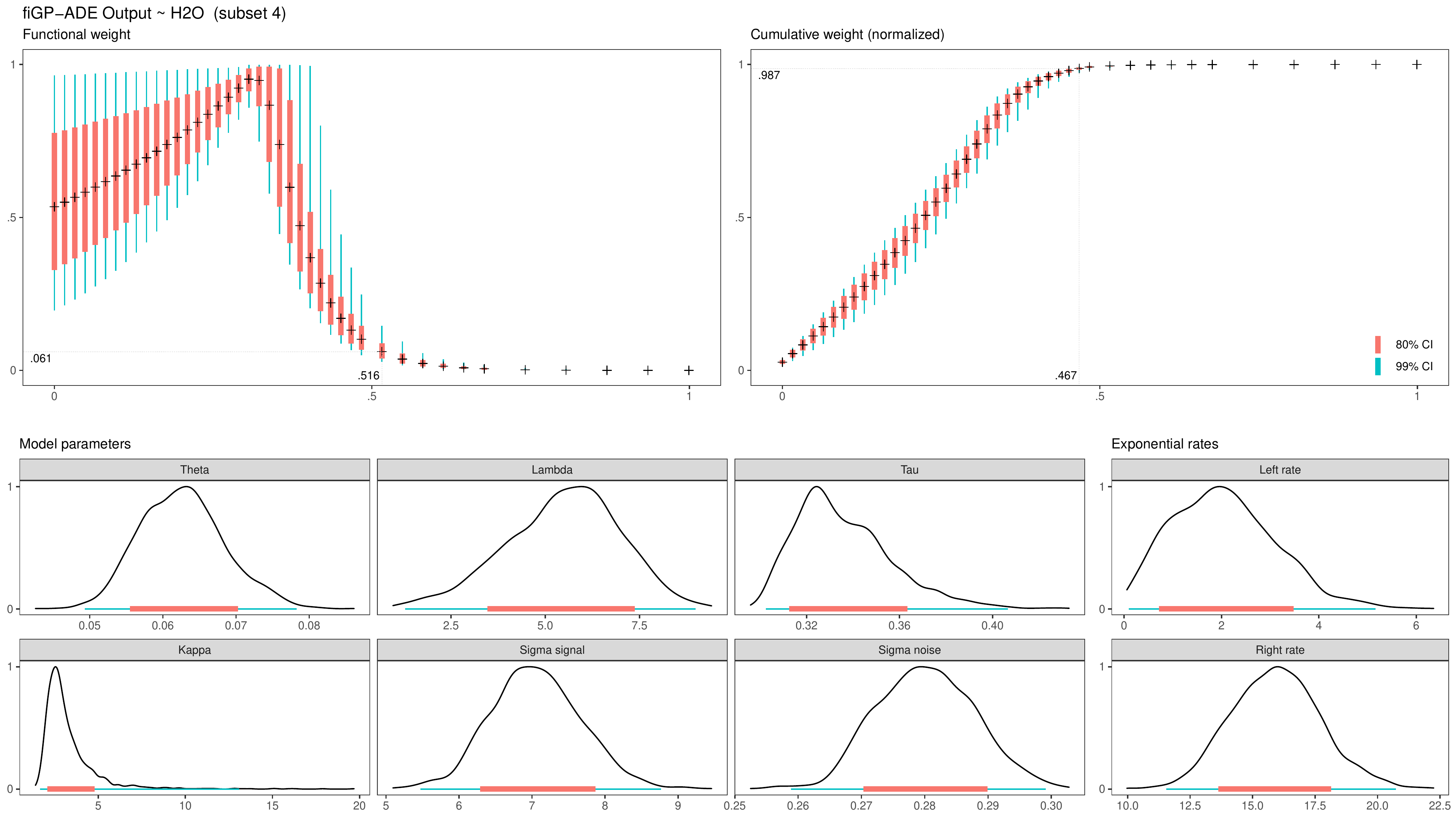}
    \caption{Estimated posterior intervals of the functional weight
      $\omega(t_k)$ (top left), $W(k) = \sum_{z = 1}^{k} \omega(t_z)
      / \sum_{z = 1}^{K} \omega(t_z) $ (top right). Model parameter
      posterior density (bottom).
    }%
    \label{fig:param-ADE-H2O}
  \end{figure}

  \clearpage

  \section*{Parameter posterior distributions (cont'd)}%
  \begin{figure}[H]
    \includegraphics[width=\linewidth]{%
      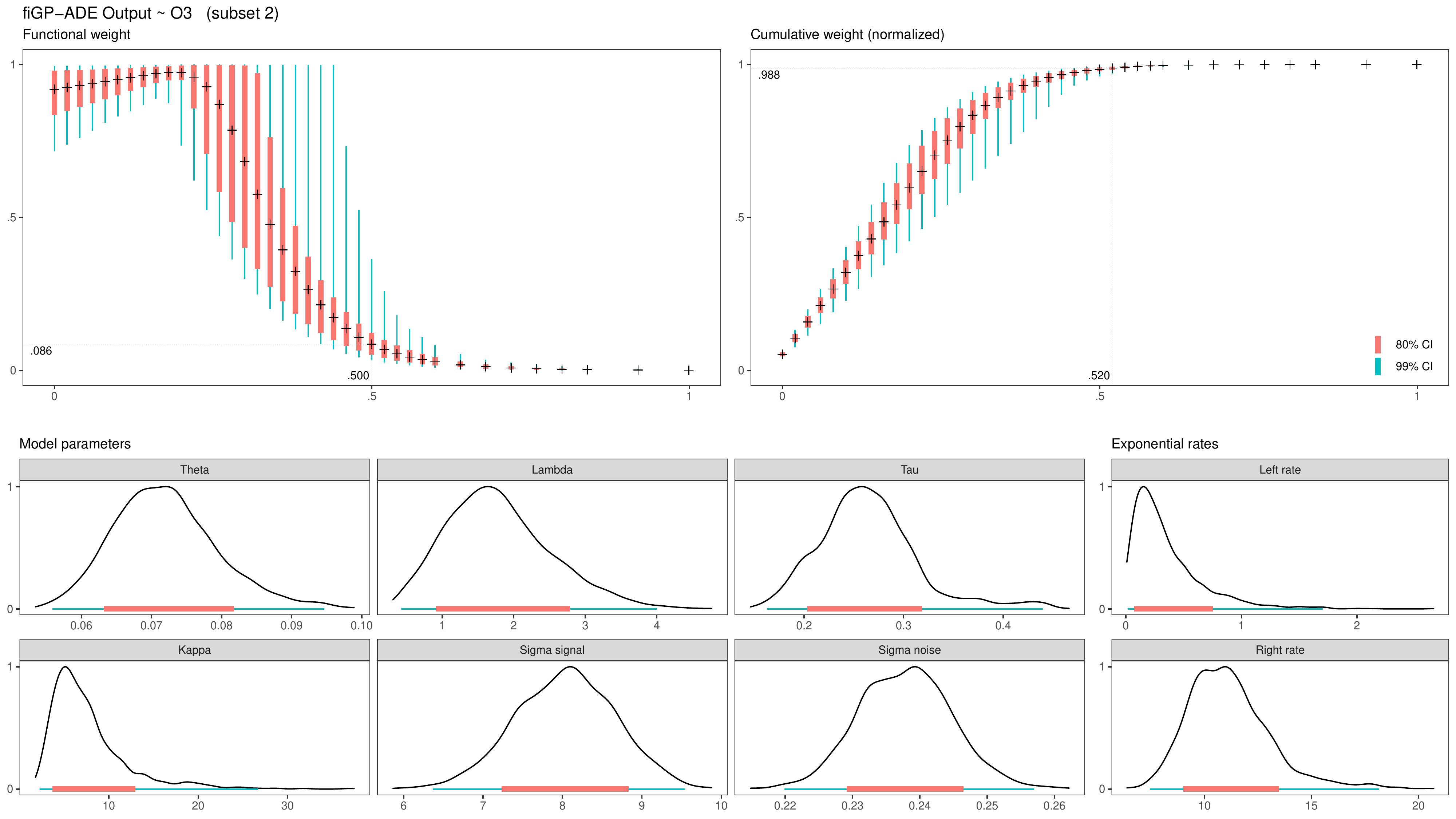}
    \caption{Estimated posterior intervals of the functional weight
      $\omega(t_k)$ (top left), $W(k) = \sum_{z = 1}^{k} \omega(t_z)
      / \sum_{z = 1}^{K} \omega(t_z) $ (top right). Model parameter
      posterior density (bottom).
    }%
    \label{fig:param-ADE-O3}
  \end{figure}

  \clearpage

  \section{Model predictions}%
  \begin{figure}[H]
    \centering
    \includegraphics[width=.7\linewidth]{%
      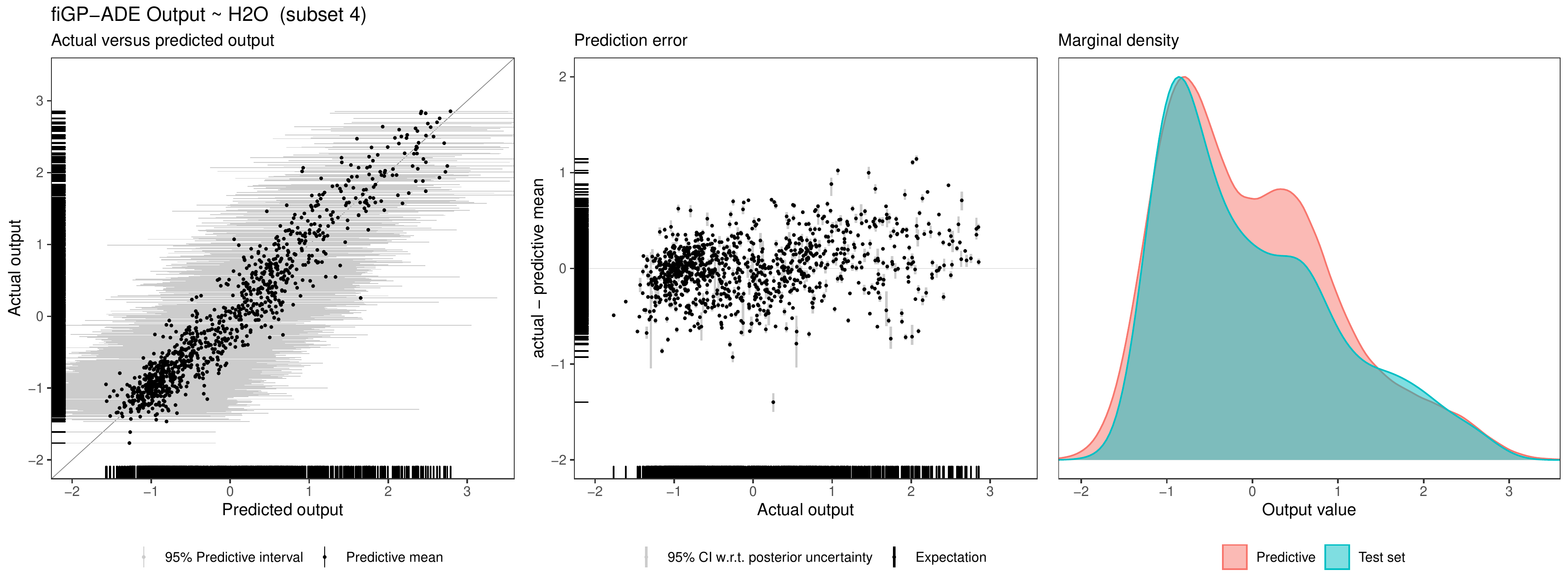}
    \includegraphics[width=.7\linewidth]{%
      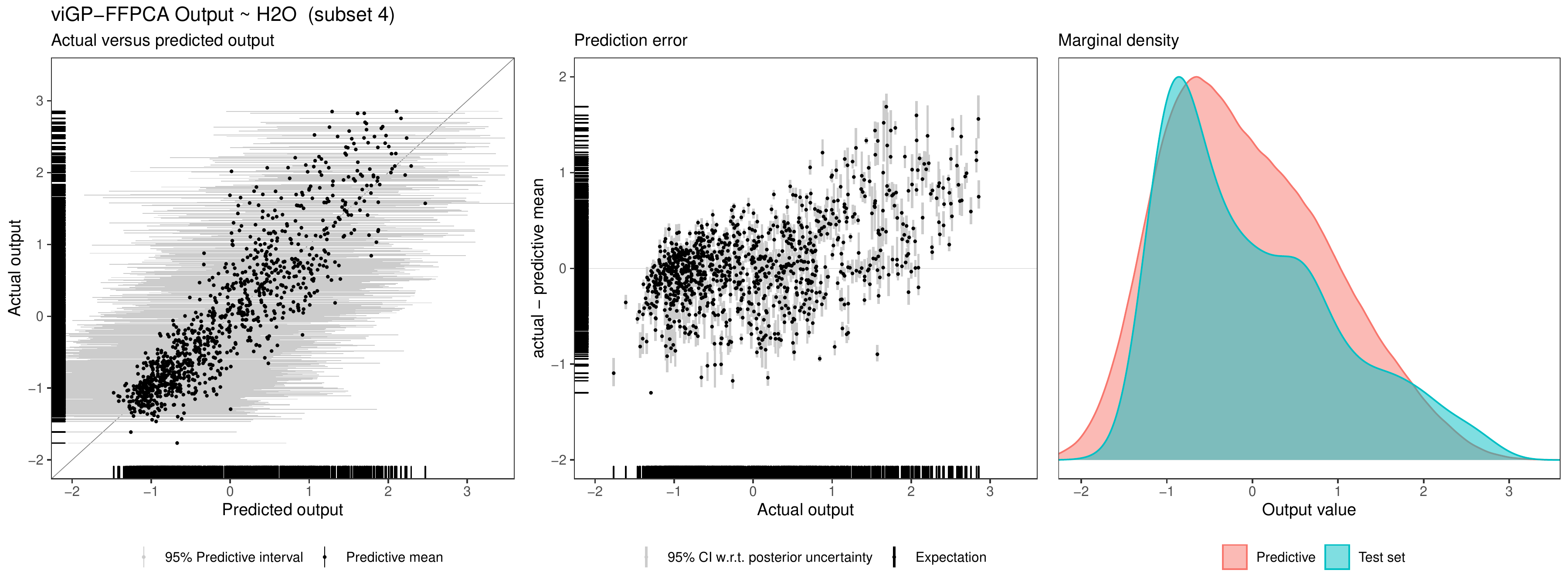}
    \caption{Out of sample output prediction for radiance as a
      function of H2O using the \textsc{ADE} (top) and the \textsc{FFPCA}
      (bottom) models.}%
  \label{fig:pred-ADE-H2O}
  \end{figure}

  \clearpage

  \section*{Model predictions (cont'd)}%
  \begin{figure}[H]
    \centering
    \includegraphics[width=.7\linewidth]{%
      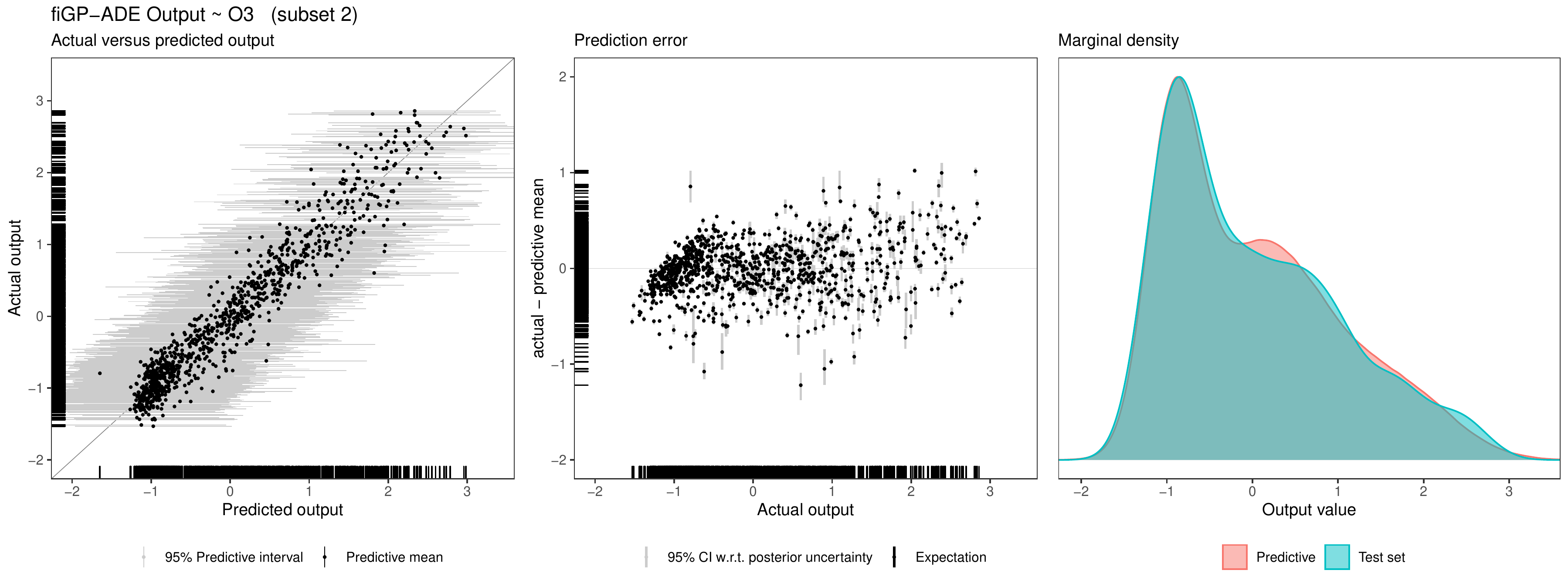}
    \includegraphics[width=.7\linewidth]{%
      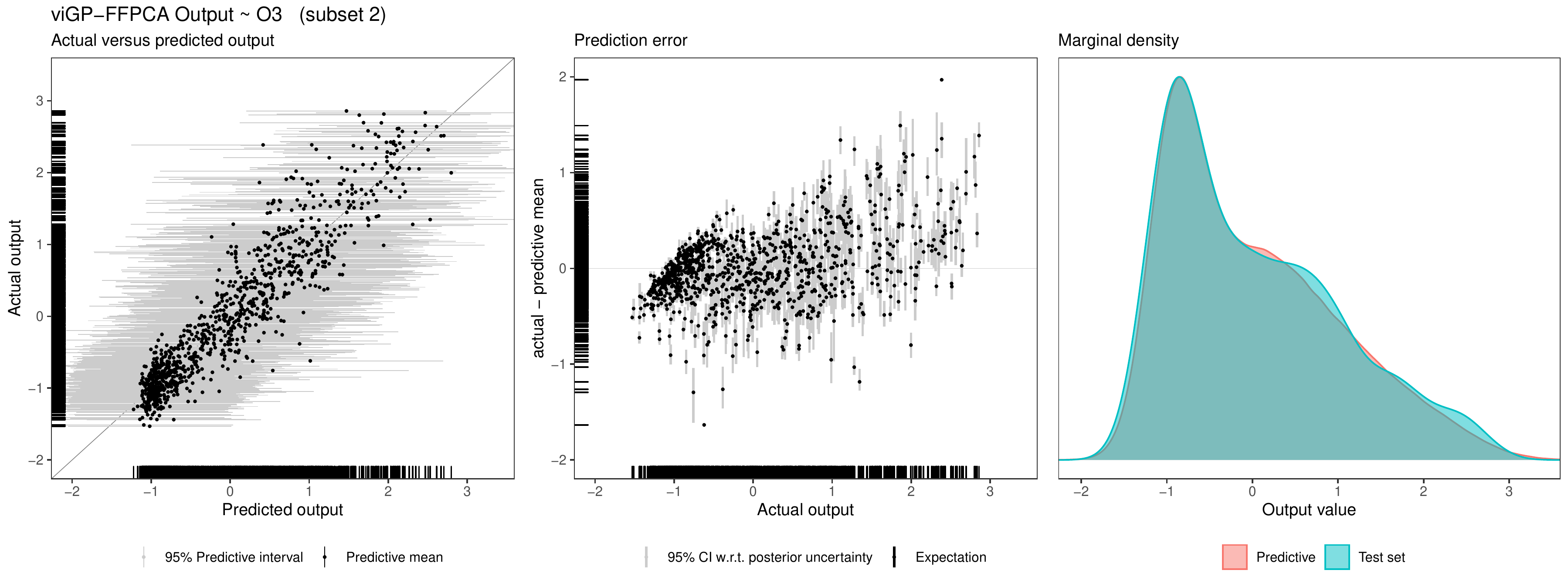}
    \caption{Out of sample output prediction for radiance as a
      function of O3 using the \textsc{ADE} (top) and the \textsc{FFPCA}
      (bottom) models.}%
    \label{fig:pred-ADE-O3}
  \end{figure}

  \clearpage

\end{landscape}


}
